\documentclass[%
superscriptaddress,
twocolumn,
 amsmath,amssymb,
 aps,
]{revtex4-2}

\usepackage{graphicx}
\usepackage{dcolumn}
\usepackage{bm}
\usepackage{amsmath,amssymb,amsthm,bm,mathtools,amsfonts,mathrsfs,bbm,dsfont} 
\usepackage{graphicx}
\usepackage{braket}
\usepackage{color}
\usepackage{float}
\usepackage[skins,theorems]{tcolorbox}
\usepackage{soul}
\usepackage[paperwidth=210mm,paperheight=297mm,centering,hmargin=1.5cm,vmargin=3.5cm]{geometry}

\usepackage[unicode=true, breaklinks=false, pdfborder={0 0 1}, backref=false, colorlinks=true, linkcolor=blue, citecolor=blue, urlcolor=black]{hyperref}

\renewcommand{\vec}[1]{\boldsymbol{#1}}

\newcommand{\vd}{\vec{d}}
\newcommand{\vw}{\vec{w}}
\newcommand{\vsigma}{\vec{\sigma}}
\newcommand{\vbeta}{\vec{\beta}}

\newcommand{\la}{\langle}
\newcommand{\ra}{\rangle}

\newcommand{\Order}{\mathcal{O}}
\newcommand{\cL}{\mathcal{L}}

\newcommand{\da}{\dagger}
\newcommand{\eps}{\varepsilon}

\newcommand{\Op}[1]{\hat{#1}}

\newcommand{\oa}{\Op{a}}
\newcommand{\ob}{\Op{b}}

\newcommand{\oH}{\Op{H}}
\newcommand{\oD}{\Op{D}}

\newcommand{\oB}{\Op{B}}
\newcommand{\oU}{\Op{U}}

\newcommand{\oC}{\Op{C}}
\renewcommand{\oc}{\Op{c}}

\newcommand{\id}{\ensuremath{\mathbbm 1}}

\newcommand{\diff}{\mathrm{d}}

\newcommand{\Tr}{\textrm{Tr}}
\newcommand{\vac}{\textrm{vac}}

\begin{document}

\title{Cascaded Optomechanical Sensing for Small Signals }

\author{Marta Maria Marchese}
\affiliation{Institute of Science and Technology Austria (ISTA), 3400 Klosterneuburg, Austria}

\author{Daniel Braun}
\affiliation{Institute for Theoretical Physics, Eberhard Karls Universität Tübingen, Auf der Morgenstelle 14, 72076 Tübingen, Germany}

\author{Stefan Nimmrichter}
\thanks{These authors share last authorship\\
marta.marchese@ist.ac.at\\
daniel.braun@uni-tuebingen.de\\
dennis.ratzel@ucl.ac.uk\\
stefan.nimmrichter@uni-siegen.de}
\affiliation{Naturwissenschaftlich-Technische Fakult\"{a}t, Universit\"{a}t Siegen, Walter-Flex-Stra\ss e 3, 57068 Siegen, Germany}

\author{Dennis R\"atzel}
\thanks{These authors share last authorship\\
marta.marchese@ist.ac.at\\
daniel.braun@uni-tuebingen.de\\
dennis.ratzel@ucl.ac.uk\\
stefan.nimmrichter@uni-siegen.de}
\affiliation{Department of Physics and Astronomy, University College London, Gower Street, WC1E 6BT London, United Kingdom}
\affiliation{ZARM, University of Bremen, Am Fallturm 2, 28359 Bremen, Germany}

\begin{abstract}
We propose a sensing scheme for detecting weak forces that achieves sensitivity scaling proportional to the number of individual sensors without relying on entanglement or other non-classical resources. Our scheme utilizes coherent averaging across a chain of $N$ optomechanical cavities, unidirectionally coupled via a laser beam. As the beam passes through the cavities, it accumulates phase shifts induced by a common external force acting on the mechanical elements. Remarkably, this fully classical approach achieves the sensitivity scaling typically associated with quantum-enhanced protocols, providing a robust and experimentally feasible route to precision sensing. Potential applications range from high-sensitivity gravitational field measurements at the Large Hadron Collider to probing dark matter interactions and detecting gravitational waves. This work opens a new pathway for leveraging coherent light-matter interactions 
for force sensing.
\end{abstract}

\maketitle

\section{Introduction}
Detecting extremely weak forces with high sensitivity remains a central challenge in modern physics, with broad relevance to both fundamental research and technological development. In fundamental studies, enhanced sensitivity enables the exploration of new physical regimes, including the investigation of gravity at the quantum scale, the search for dark matter candidates, and the examination of the principles underlying quantum mechanics. From a technological perspective, advances in ultra-sensitive force detection underpin the development of precision measurement techniques and sensor technologies with wide-ranging applications.

Optomechanical systems~\cite{aspelmeyer2014cavity} have long been used to tackle diverse questions in quantum mechanics, from the exploration of fundamental physics~\cite{grossardt2016optomechanical,bahrami2014proposal,carney2021ultralight} to sensing applications~\cite{nimmrichter2014optomechanical,bose2025massive,braun2025metrology}. The ground state cooling of the mechanical element and the radiation-pressure interaction with a light field allow to enter a quantum regime where the mechanical motion becomes highly sensitive to small disturbances~\cite{qvarfort2018gravimetry,schneiter2020optimal,qvarfort2021optimal}. In this context, optomechanical systems represent ideal platforms to test weak forces as they offer high force sensitivity with many recent advances~\cite{kippenberg2008cavity,iwasawa2013quantum,tsang2013quantum}. 
This opens new paths to unveil and test physical effects, ranging from quantum-gravity signatures~\cite{bose2017spin,marletto2017gravitationally,krisnanda2020observable,carlesso2019testing,miki2024quantum}, to dark matter detection~\cite{monteiro2020search,carney2021mechanical,brady2023entanglement,qin2025mechanical}, or measurements of high-frequency gravitational waves from continuous sources~\cite{arvanitaki2013detecting}. 

Improving a detector’s sensitivity means making the signal stand out more clearly from background noise. A common method is to average the results over $N$ measurements, either by repeating the same experiment or by collecting data from multiple independent probes. This approach leads to a Signal-to-Noise Ratio (SNR) enhanced by a factor proportional to $\sqrt{N}$, known as the Standard Quantum Limit (SQL). Quantum correlations between probes can push this improvement even further, reaching what is known as the Heisenberg limit (HL), where the SNR scales linearly with $N$, the number of probes~\cite{giovannetti2004quantum}. Although entanglement and squeezing have demonstrated such gains, their practical use is limited by the fragility of quantum resources under the effects of decoherence, which significantly restrict the number of usable probes in an experiment ~\cite{leibfried2005creation,nagata2007beating}. 

Another possibility to surpass the standard quantum limit is to couple the $N$ probes simultaneously to a single quantum system (``bus") and eventually read out the latter instead of measuring the probes individually. In the literature, the term coherent averaging is used for this approach~\cite{fraisse2015coherent,braun2018quantum,braun_coherently_2014}.
The phase information from the $N$ probes accumulates in the quantum bus, leading to an improvement of the SNR without entanglement or quantum resources; even simple product states of the probes are sufficient, and in certain regimes they can achieve full HL scaling. Because it avoids fragile entangled states, the scheme can be more robust to decoherence, making it significantly more practical in realistic, noisy experimental settings. Among the several other ways that have been found to bypass the entanglement requirement to reach HL sensitivity, we just mention non-linear schemes~\cite{boixo2007generalized,beltran2005breaking}, multi-pass schemes~\cite{higgins2007entanglement}, or non-unitary encoding of the parameter~\cite{marzolino2013precision}.
In fact, coherent averaging can be seen as an interference phenomenon that also works in the classical wave regime, where it is used for SNR enhancement in multipass microscopy of sensitive samples \cite{juffmann2016multi,nimmrichter2018full}. In \cite{braun_coherently_2014} it was suggested for enhancing the sensitivity of measurements of Newton's constant $G$ using a set of classical harmonic oscillators all coupled to a central one. 
Alternatively, a classical coherent averaging of measurement signals can also enhance the sensitivity of optomechanical sensor arrays: instead of a single quantum bus readout of the accumulated output signal, one performs individual homodyne measurements on the sensors, all driven by the same continuous laser field, and combines the data in post processing based on fixed temporal relations between the readout signals \cite{brady2023entanglement}.

Here, we propose a sensing scheme that exploits coherent averaging for cascaded optomechanical systems. The systems are connected through a unidirectional optical field that acts as the bus and accumulates all the phase signals before being measured. As the readout is performed on the same light mode, the necessary temporal relation between different light-sensor interactions is fixed by adjusting the setup geometry. In contrast to parallel homodyne measurements where individual path length have to be adjusted on the scale of the optical wave length, the necessary precision in the cascaded geometry is defined by the mechanical frequency as only the accumulating phase shifts linked to the oscillator phases of the mechanical elements have to be matched. This is a practical advantage while the accumulation of loss in the cascaded geometry imposes additional constraints on each sensor system as well as on the routing of light through the cascade.  

We investigate three different readout regimes: (a) the stroboscopic regime with light pulses much shorter than the mechanical period, (b) the continuous-wave regime with light pulses that are longer than the mechanical period and the signal duration, and (c) the continuous wave regime with light pulses that are shorter than the signal duration. For all three regimes, we find that Heisenberg-like scaling can be achieved in the ideal case without photon losses.
We show that in presence of losses, even though the $N^2$-scaling is lost, there is an enhancement up to an optimal number of systems $N_{\rm opt}$ in comparison to the standard quantum limit. To show the utility of the scheme, we discuss three example applications: (1) detection of gravitational field at LHC, (2) dark matter detection, (3) gravitational waves detection.

The proposed configuration is conceptually analogous to sequential multi-pass schemes, in which precision improves by repeatedly sending light through the same cavity. However,  the cascaded system offers a complementary approach in which multiple sensors independently receive the same signal. This may be advantageous in scenarios where a single system cannot be reused, such as when a high-intensity laser pulse induces incoherent absorption, loss, or damage.

We note that the cascaded setup inherently introduces a time delay between consecutive cavities due to light propagation. In other words, the bus receives each probe signal sequentially, seemingly at odds with the assumption of simultaneous probe–bus couplings for coherent averaging. This however depends on the timing of the probe signals.
For example, the signal may be imprinted on consecutive mechanical elements with a time delay and the light propagation may be adjusted to exactly match this delay. This can be the case in one of our example applications, the detection of the gravitational field at the LHC. Furthermore, if decoherence of the mechanical elements is happening on a much longer time scale than their oscillation period, simultaneity can effectively be achieved by setting the time delay between two oscillators to match an integer multiple of the mechanical period.
On the other hand, should the total delay across the chain 
remain negligible compared to the mechanical period, the interactions are rendered effectively simultaneous.


In Sec.~\ref{sec:system} we describe the cascaded optomechanical setup and how the signal is imprinted on the motion of the mechanical element in each cavity; by deriving and iterating the input-output relations, we obtain a general recursive formula for the output field amplitude. The solution to this recursive relation is given explicitly in the following three subsections: in ~\ref{subsec:stroboscopic} for the stroboscopic regime, in ~\ref{subsec:continwave} for the continuous-wave regime, and in~\ref{subsec:continwavecontinsignal} for the continuous-wave regime with a continuous signal. In Sec.~\ref{sec:metrology} we introduce the metrological bounds to the signal-to-noise ratio (SNR) for Gaussian states in the weak coupling regime. The results for the three limiting cases are presented in Sec.~\ref{subsec:SNRresults}, and the upper bound case obtained when all the cavities are identical in Sec.~\ref{subsec:equalcav}. In Sec.~\ref{sec:applications} we discuss three potential applications of the cascaded sensing scheme: ~\ref{subsec:DM} Dark Matter searches, ~\ref{subsec:GW} Gravitational waves, and ~\ref{subsec:LHC} Gravitational field of ultra-relativistic matter. In Sec.~\ref{sec:conclusions} we draw our conclusions and outline possible directions for future work. All parameters used in the paper are listed in Table~\ref{tab:parameters}.

\begin{figure*}
\includegraphics[width=0.8\textwidth]{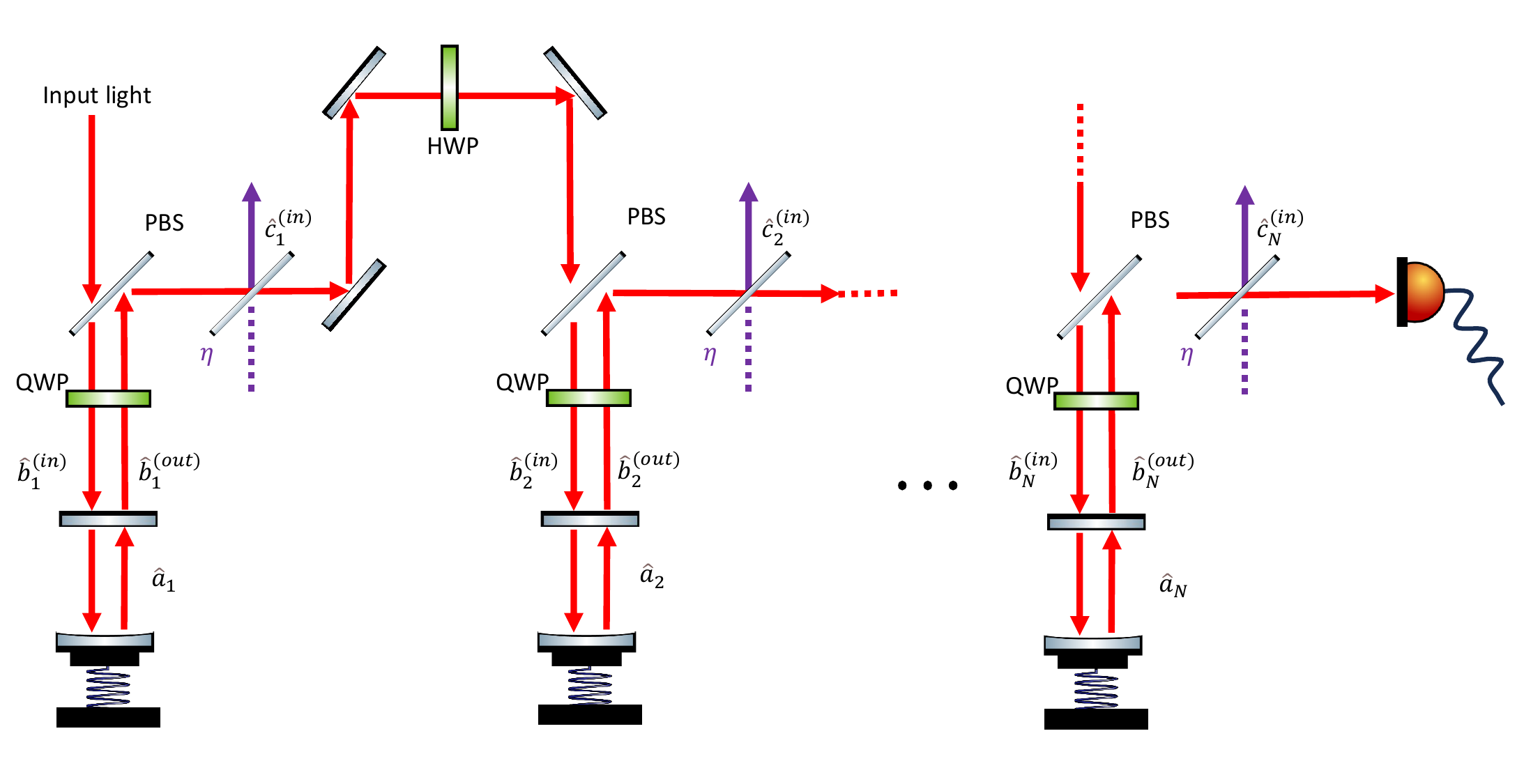}
\caption{Sketch of the cascaded system with $N$  optomechanical cavities. An input optical pulse enters the first cavity and, using suitable optical elements (quarter-wave plates (QWPs) and polarizing beam splitters (PBSs)), is directed through all subsequent cavities before reaching the detector. During its propagation, the pulse interacts with each mechanical resonator, acquiring a phase shift that cumulatively encodes the contributions from all $N$ systems. }
\label{fig:scheme}
\end{figure*}

\section{Cascaded Optomechanics}\label{sec:system}

We implement a coherent-averaging scheme to detect weak signals encoded as phase shifts in mechanical resonators. The system, shown in Fig.~\ref{fig:scheme}, consists of $N$ optomechanical cavities, acting as parameter-dependent probes. They are uni-directionally coupled through laser light that serves as a common ``quantum bus''. By fixing the initial polarization and inserting 
{polarizing beam splitter (PBS) and quarter wave plates (QWP) (or alternatively Faraday rotators)}
along the path, it is possible to drive the pulse through all the cavities avoiding back reflected light. As the optical pulse enters each cavity, it interacts with the respective mechanical resonator. An external signal that perturbs the resonator motion
imprints a phase shift $\varphi$ on the bus field. These shifts accumulate coherently in the input laser pulse, amplifying the effect of the signal. The output pulse is eventually measured to extract information on the phase shift induced by the signal on the mechanical motion.

\begin{table}[t]
\centering
\begin{tabular}{lc}
\hline
Meaning &
Symbol  \\
\hline 
Frequency of $n^{\rm th}$ mechanical resonator & $\Omega_n$ \\
Bandwidth of the $n^{\rm th}$ mechanical signal & $\Delta \Omega_n$ \\
Standard deviation of mechanical phase fluctuations & $\sigma_\phi$ \\
Frequency of $n^{\rm th}$ cavity & $\omega^c_n$ \\
Detuning of $n^{\rm th}$ cavity & $\Delta_n$ \\
Linewidth of $n^{\rm th}$ cavity & $\kappa_n$\\
Coupling rate of $n^{\rm th}$ optomechanical system & $g_n$\\
Probe pulse duration & $\tau$ \\
Probe light frequency & $\omega_L$ \\
Attenuation factor quantifying optical loss per cavity & $\eta$ \\
Inter-cavity delay & $T$ \\
Total number of optomechanical systems & $N$ \\
Number of input photons & $N_{\rm in}$\\
\hline
\end{tabular}
\caption{\label{tab:parameters} Parameters used in this article.}
\end{table}

We model each optomechanical system $n$ as a single optical cavity mode with frequency $\omega_n$ that is modulated proportionally to the position $X_n(t)$ of a classical end mirror oscillating with central frequency $\Omega_n$. We assume that all cavity modes are initially in the ground state $|0\rangle $. The signal is assumed to be imprinted on the motion of the end mirrors before the start of the measurement.
The dynamics of all light modes (cavity and probe light) is expressed in the reference frame of the (central) frequency of the probe light $\omega_L$. Hence, for each cavity, we have the Hamiltonian
\begin{equation}
    \begin{aligned}
    \hat H_n&= \hat H_{n,a,0}+ \hat H_{n,b,0} +\hat H_{n,I}\,,\,\,\text{where} \\
\hat H_{n,a,0} &= \hbar(\Delta_n - g_n q_n(t))\hat a_n^\dagger \hat a_n,  \\
\hat H_{n,b,0} &= \hbar\int d\omega\;(\omega-\omega_L) \hat b_n^\dagger (\omega) \hat b_n(\omega), \\
\hat H_{n,I} &= i \hbar  \sqrt{\frac{\kappa_n}{2\pi}} \int d\omega \left(\hat b_n^\dagger(\omega)\hat a_n-\hat b_n(\omega)\hat a^\dagger_n\right).
    \end{aligned}
\label{eq:hamiltonian}
\end{equation}
$\hat{H}_{n,a,0}$ describes the single-mode field of the $n^{\text{th}}$ cavity with detuning  $\Delta_n=\omega_n^c-\omega_L$, expressed in terms of bosonic annihilation and creation operators, which satisfy the standard commutation relation  $[\hat{a}_n,\hat{a}_n^\dagger]=1$. The term also contains the shift of the cavity frequency due to the change in position of the mechanical element: the parameter $g_n$ denotes the optomechanical coupling between the mechanical resonator and the cavity field, while $q_n(t)=X_n(t)\sqrt{2m_n\Omega_n/\hbar}$. The second Hamiltonian term $\hat{H}_{n,0,b}$ corresponds to the free Hamiltonian of the outside field, representing a mode continuum in terms of operators  $\ob^\dagger_n(\omega)$ and $\ob_n(\omega)$, which satisfy the canonical commutation relations  $[\hat{b}_n(\omega),\hat{b}_n^\dagger(\omega')]=\delta(\omega-\omega')$. Finally, the last term $\hat{H}_{n,I}$ models the interaction at the fixed mirror between the cavity and input fields as a beam-splitter with frequency-independent coupling given by the linewidth $\kappa_n$ (Markovian approximation~\cite{gardiner2017quantum}). This form of the coupling is a good approximation for high-finesse cavities, that is, when the cavity roundtrip time $\tau_{\rm rt}$ is much shorter than the time scale of cavity decay (see Chapter 9 of \cite{dutra2005cavity}). Therefore, we make the assumption $\kappa_n \ll 1/\tau_{\rm rt}$.
While many of the following steps are general, eventually, we shall operate in the bad cavity regime in which $\Omega_n \ll \kappa_n$. The optomechanical coupling rates $g_n$ are later also assumed small. The general restrictions used throughout the whole analysis are listed in Table~\ref{tab:generalapproximations}.\\

\begin{table}[t]
\centering
\label{tab:generalapproximations}
\small
\begin{tabular}{p{0.56\columnwidth} p{0.20\columnwidth} 
}
Frequency-independent coupling &
$\kappa_n(\omega)\approx\kappa_n$
\\

High-finesse cavity &
$\tau_{rt}\ll 1/\kappa_n$
\\

Bad-cavity regime &
$\Omega_n\ll\kappa_n$
\\

Weak coupling regime &

$g_n\ll\kappa_n$  
\\

\hline
\end{tabular}
\caption{General restrictions of the considered regime. In addition, we approximate the effect of optical losses as an attenuation of the light field amplitude.}
\end{table}

The dynamical degrees of freedom for each cavity are the intra-cavity field $\hat{a}_n$ and the external field $\hat{b}_n(\omega)$. Their dynamics is governed by the Heisenberg equations of motion. Following the standard approach \cite{gardiner1985input} (see Appendix~\ref{app:recursion}), we can obtain a formal solution for $\ob_n (\omega,t)$ which we substitute into the equation of motion for $\oa_n (t)$ to obtain the cavity Langevin equation,
\begin{equation}\label{eq:an_dt_main}
    \partial_t \oa_n (t) = \left[-\Gamma_n +ig_nq_n(t) \right] \oa_n(t) - \sqrt{\kappa_n} \ob_n^{\rm (in)} (t),
\end{equation}
where we used the abbreviation $\Gamma_n = \kappa_n/2 + i\Delta_n$. The input field is defined as 
\begin{equation}
    \ob_n^{\rm (in)}(t) = \int \frac{\diff \omega}{\sqrt{2\pi}}e^{-i(\omega-\omega_L)(t-t_0)} \ob_n (\omega,t_0),
    \label{eq:INPUT}
\end{equation}
and $t_0 \to -\infty$ is the asymptotic initial time at which the cavity and the light field are still decoupled, $[\oa_n(t_0),\ob_n^\da (\omega,t_0)]=0$. With $t_1\to \infty$, we can define an analogous expression for the output field, 
\begin{equation}
    \ob_n^{\rm (out)}(t) =  \int\frac{\diff \omega}{\sqrt{2\pi}} \, e^{-i(\omega-\omega_L)(t-t_1)} \ob_n (\omega;t_1),
    \label{eq:OUTPUT}
\end{equation}
where $\ob_n(\omega,t_1)$ is the free field for $t_1\rightarrow+\infty$, and thereby obtain the input-output relation ~\cite{gardiner1985input},
\begin{equation}\label{eq:IO_bn_main}
    \ob_n^{\rm (out)} (t) = \ob_n^{\rm (in)} (t) + \sqrt{\kappa_n} \oa_n (t).
\end{equation}
As the state of the input light field, we consider a Gaussian laser pulse of length $\tau$
\begin{equation}\label{eq:psi_in_main}
    |\psi (t_0)\ra = \prod_\omega \oD_1 \left[   
    \bar\beta \tau 
    e^{-i\omega t_0 - \tau^2(\omega-\omega_L)^2/2} \right] |\vac\ra,
\end{equation}
where the continuous wave regime corresponds to $\tau\to\infty$.

In our cascaded setup, the input field of the $n^{\mathrm{th}}$ cavity $b_{n}^{\rm (in)}(t)$ is determined by the output field of the preceding cavity, subject to a propagation delay $T$. We model possible losses between the cavities by beam-splitter transformations that remove a fraction of the output field and inject an equivalent amount of vacuum noise into the next input field represented by vacuum fields $\oc_{n}^{\rm (in)} (t)$. This gives rise to the condition
\begin{equation}\label{eq:OI_n_np1_loss_main}
 \ob_n^{\rm (in)} (t) = \ob_{n-1}^{\rm (out)} (t-T)e^{i\chi} + \oc_{n-1}^{\rm (in)}(t-T) e^{i\xi },
\end{equation}
with the abbreviations $\chi = \omega_L T-i\ln \sqrt{\eta}$, $\xi = \omega_L T - i\ln \sqrt{1-\eta}$, and the attenuation parameter $\eta \in [0,1]$. Note that we shift the vacuum fields in time and append a phase for convenience. The above formula applies for $n\geq 2$ as we do not account for extra losses before the first cavity.
Iterating \eqref{eq:IO_bn_main} and \eqref{eq:OI_n_np1_loss_main}, the $n$-th output field can be expressed in terms of the original first input field, the cumulative response of all $n$ cavities, and the cumulative vacuum noise input,
\begin{align}
    \ob_n^{\rm (out)} (t) &= e^{i(n-1)\chi} \Big[ \ob_1^{\rm (in)}(t-(n-1)T) \\
    & \quad + \oB_n (t-(n-1)T)+\oC_n(t-(n-1)T)\Big] ,\nonumber 
\end{align}
introducing the cumulative operators,
\begin{equation}
    \begin{aligned}
        \label{eq:Bn_def_loss}
\oB_n (t) &=  \sum_{k=1}^{n}\sqrt{\kappa_k} \oa_k (t+(k-1)T) e^{-i(k-1)\chi}, \\
\oC_n (t) &= e^{i\xi}\sum_{k=1}^{n-1}\oc_k^{(\rm in)} (t+(k-1)T) e^{-ik\chi}.
    \end{aligned}
\end{equation}
Ideally, the mechanical signal should coincide with the arrival time of the pulse at the $n$-th cavity, $(n-1)T$. Let us therefore express the mechanical amplitude as $q_n(t) = Q_n (t-(n-1)T)$, with $Q_n (t)$ ideally centered around $t=0$. Then, we integrate the cavity Langevin equation \eqref{eq:an_dt_main} for $t> t_0$, make use of the bad cavity regime and let $t_0 \to -\infty$ to obtain the recursive relation (see Appendix \ref{app:recursion})
\begin{align}
            \oB_n (t) &= \oB_{n-1} (t) -\kappa_n \int_{0}^{\infty} \diff t' \, e^{-G_n(t)t'} \Big[ \ob_1^{\rm (in)} (t-t') \nonumber \\
            & \quad + \oB_{n-1}(t-t') + \oC_n(t-t')\Big].
    \label{eq:Bn_recursion_loss_main}
\end{align}
where $\oB_0 \equiv 0$ and $\oC_1 \equiv 0$ and $G_n(t) := \Gamma_n - ig_n Q_n (t)$. We will be mainly concerned with the mean value of the cumulative response of $N> 1$ cavities, $\la \oB_N (t)\ra$, as it is the coherent output field amplitude by which we measure the mechanical signal. When taking the expectation value, the vacuum noise input always vanishes, $\la \oC_n \ra = 0$. Therefore, the recursion formula~\eqref{eq:Bn_recursion_loss_main} can be formulated in terms of the time-shifted output amplitude, $\beta_n(t) = \beta(t) + \la \oB_n(t)\ra$ as
\begin{equation}\label{eq:betan_recursion_main}
    \beta_n(t) = \beta_{n-1} (t) - \kappa_n \int_{0}^{\infty} \diff t' \, e^{-G_n(t)t'} \beta_{n-1} (t-t'),
\end{equation}
where $\beta_0(t)=\beta(t):=\la \ob_1^{\rm (in)} (t)\ra=\bar\beta e^{-t^2/2\tau^2}$
and 
\begin{equation}\label{eq:betan_bout_loss_main}
    \la \ob_n^{\rm (out)} (t)\ra = \eta^{(n-1)/2}\beta_n (t-(n-1)T)e^{i(n-1)\omega_L T}.
\end{equation}
We notice that the attenuation $\chi$ does not explicitly appear in the recursion formula~\eqref{eq:Bn_recursion_loss_main}. The attenuation only enters in the dependence of the actual output field amplitude after $n$ cavities on the $\beta_n$ in Eq.~\eqref{eq:betan_bout_loss_main}, by the $(n-1)$-fold attenuation factor.

We denote the Fourier transform of the time-shifted output amplitude as $\tilde\beta_n(\omega) = \int \diff t \, e^{i\omega t}\beta_n(t)/\sqrt{2\pi}$ and use the inverse Fourier transform to translate the recursion formula \eqref{eq:betan_recursion_main} into frequency space,
\begin{align}\label{eq:recursion-beta-F}
\tilde\beta_n(\omega) &=  \tilde\beta_{n-1}(\omega) - \frac{\kappa_n}{2\pi} \int\diff\omega' \tilde\beta_{n-1}(\omega') \underbrace{\int\diff t \, \frac{e^{i (\omega-\omega')t}}{G_n (t)-i\omega'}}_{=: F_n (\omega',\omega-\omega')} ,
\end{align} 
where $\tilde\beta_0(\omega)=\tilde\beta(\omega) = e^{i\omega t_0} \la \ob_1^{\rm (in)} (\omega+\omega_L)\ra=\bar\beta \tau 
    e^{- \tau^2\omega^2/2}$.
As we are concerned with the measurement
of weak mechanical signals and realistic optomechanical coupling rates are typically small, $g_n \ll \kappa_n$, we can reasonably expand to lowest order in $|g_n Q_n(t)|/\kappa_n$,
\begin{equation}
    \begin{aligned}
    F_n(\omega',\Omega) &\approx \frac{2\pi \delta(\Omega)}{\Gamma_n - i\omega'} + \frac{\sqrt{2\pi}ig_n \tilde Q_n(\Omega)}{(\Gamma_n - i\omega')^2},
    \end{aligned}
\end{equation}
where $\tilde Q_n(\Omega) = \tilde q_n (\Omega) e^{-i\Omega (n-1)T}$ is the Fourier transform of the mechanical displacement. 
For convenience, we define the integral operator $\cL_n$ acting on a frequency-space function by
\begin{equation}\label{eq:Ln_main}
    \cL_n [\tilde\alpha](\omega) := \int\diff\omega' \left[1 - e^{i\phi_n(\omega')} \right]^2 \tilde\alpha (\omega') \frac{\tilde Q_n(\omega-\omega') }{\sqrt{2\pi}},
\end{equation}
with the cavity response phase $e^{i\phi_n(\omega)} = -(\Gamma_n-i\omega)^*/(\Gamma_n - i\omega)$, i.e., $\phi_n(\omega)=\pi + 2\arctan(2(\omega-\Delta_n)/\kappa_n)$. The recursion~\eqref{eq:recursion-beta-F} then reads
\begin{equation}
    \tilde\beta_n = e^{i\phi_n} \tilde\beta_{n-1} - i\eps_n \cL_n [\tilde\beta_{n-1}], \qquad \tilde\beta_0 = \tilde\beta,
\end{equation}
noting that all $\tilde\beta_k$ and $\phi_k$ are functions of $\omega$ and $\eps_n=g_n/\kappa_n$. To first order in $\eps_n$, the consistent solution to this recursion for $N$ cavities can be stated as
\begin{equation}\label{eq:solution_betan_main}
    \tilde\beta_n \approx e^{i\sum_{j=1}^N \phi_j} \tilde\beta - i\sum_{k=1}^N \eps_k e^{i\sum_{j=k+1}^N \phi_j} \cL_k \left[ e^{i\sum_{\ell=1}^{k-1} \phi_{\ell} }\tilde\beta \right], 
\end{equation}
using the convention $\sum_{N+1}^N (\ldots) \equiv 0$. In general, the result cannot be given analytically, but there are two limiting regimes in which approximate analytic treatments are possible: the stroboscopic regime of short light pulses compared to the mechanical signal and the continuous-wave (CW) regime of a light pulse that is much longer than the mechanical signal. The former regime can also be considered beyond the weak-coupling limit (see Appendix \ref{app:strobobeyondweak}).  An analysis of the validity of the three limiting cases for the purposes of this article is provided in Appendix \ref{app:checksofapprox}.

\subsection{Stroboscopic regime}\label{subsec:stroboscopic}

The stroboscopic regime corresponds to the case of an input pulse that is short compared to the timescale of mechanical motion, $\Omega_n \tau \ll 1$,~\cite{vanner2011pulsed,pikovski2014macroscopic}. This means that not only the average input field but also all the responses $\la \oB_n(t)\ra$ and $\beta_n (t)$ remain sharply peaked in time around $t\approx 0$ compared to the mechanical motion. Conversely, in frequency space, the responses $\tilde\beta_n$ are spectrally broad compared to the $\tilde Q_k$, and we can therefore approximate $\tilde\beta_n (\omega-\Omega) \tilde Q_k (\Omega) \approx \tilde\beta_n (\omega) \tilde Q_k (\Omega)$ for any $n\geq 0$ and $k\geq 1$. In addition, the cavity response times are also assumed short, $\kappa_n \gg \Omega_n$, so that $\phi_n (\omega-\Omega) \tilde Q_k (\Omega) \approx \phi_n (\omega) \tilde Q_k (\Omega)$. Thus the integral operator \eqref{eq:Ln_main}, applied to such spectrally broad functions, can be approximated by,
\begin{equation}\label{eq:Ln_strob}
    \cL_n [\tilde\alpha](\omega) \approx \left[1 - e^{i\phi_n(\omega)} \right]^2 \tilde\alpha (\omega) Q_n (0).
\end{equation}
The 1st-order recursion formula \eqref{eq:solution_betan_main} simplifies to,
\begin{equation}\label{eq:solution_betan_strobWeak_main}
    \begin{aligned}
            \tilde\beta_n(\omega)
            & \approx e^{i\sum_{j=1}^N \left[\phi_j(\omega) + 2\frac{g_j}{\kappa_j} Q_j(0)\left( 1 - \cos\phi_j(\omega)\right) \right]}\tilde\beta(\omega) .
    \end{aligned}
\end{equation}
We find that the signals imprint a cumulative spectral phase modulation onto the input field. This result holds, in particular, when the mechanical displacement is constant in time, as the stroboscopic regime then applies for any $\tau < \infty$ and $\kappa >0$. The $\phi_j (\omega)$ describe the responses of the undisplaced cavities, which filter and reshape light pulses that are spectrally broad compared to the cavity linewidths.

\subsection{Continuous-wave regime}\label{subsec:continwave}

In the opposite regime, the pulse is long compared not only to the cavity lifetime, $\kappa_{n}\tau \gg 1$, and the mechanical oscillation, $\Omega_n \tau \gg 1$, but also to the duration of the mechanical signal, $\Omega_n \tau > \Delta \Omega_n \tau \gg 1$. In other words, the light pulse $\tilde\beta$ is spectrally much narrower than any of the $\phi_k$ and $\tilde Q_k$, so that, whenever the integral operator \eqref{eq:Ln_main} is applied to a function including $\tilde\beta$, one can assume $\omega'\approx 0$ in the arguments of said functions,
\begin{equation}\label{eq:Lk_cw}
    \cL_k [e^{i\sum_{\ell} \phi_{\ell}}\tilde\beta](\omega) \approx 
    \frac{\kappa_k^2}{\Gamma_k^2} e^{i\sum_{\ell}\phi_{\ell} (0)}\tilde Q_k (\omega) \beta(0),
   \end{equation}
where the time-domain amplitude $\beta(0)=\int d\omega \tilde{\beta}(\omega)/\sqrt{2\pi}$ arises due to the remaining integral in Eq.~\eqref{eq:Ln_main}.
The right hand side of Eq.~\eqref{eq:Lk_cw} is peaked in frequency space around $\pm \Omega_k$. 
The delay times $T$ between the cavities can be arbitrary.

Moreover, in Eq.~\eqref{eq:solution_betan_main}, we can also make use of the fact that the $\tilde Q_k$ are sharply peaked compared to the cavity linewidths, $\Delta \Omega_k \ll \kappa_n$ for all $\kappa,n\ge1$, and approximate $\phi_n (\omega) \approx \phi_n(0)$ whenever the phases appear together with the $\tilde Q_k$. Hence, 
\begin{equation}\label{eq:solution_betan_CW_main}
    \begin{aligned}
    \tilde\beta_n (\omega) &\approx e^{i\sum_{j=1}^N \phi_j (0)} \Big[\tilde\beta(\omega) \\
    & \quad- i\beta(0)\sum_{k=1}^N \frac{g_k\kappa_k}{\Gamma_k^2}e^{-i\phi_k(0)} \tilde Q_k (\omega)\Big].
    \end{aligned}
\end{equation}
Instead of a cumulative phase imprinted on the broad carrier light pulse in the stroboscopic regime, here we have spectrally resolved sidebands that build up with each cavity and modulate the carrier in time.

\subsection{Continuous-wave regime for continuous signals}\label{subsec:continwavecontinsignal}

There is one more relevant limit that the two discussed regimes do not cover: The light pulse could be long compared to the cavity and the mechanical oscillation frequencies, but the mechanical signals could be perfectly harmonic or quasi-continuous with a longer duration than the light pulse; that is, $\kappa\tau \gg \Omega_n \tau \gg 1 \gg \Delta\Omega_k \tau$ for all $\kappa,n\ge1$. In this case, we make use of the fact that a real-valued oscillating $Q_n (t)$ will yield a Fourier transform $\tilde Q_n(\omega)$ that is sharply peaked around $\pm \Omega_n$. This yields spectrally separated sidebands around the carrier $\tilde\beta$, which is much less sharply peaked around zero. In \eqref{eq:Ln_main}, we can now reduce the integral to two terms in which we replace the argument of the $\phi_n$ and of $\tilde\beta$ by $\omega' \approx \omega \pm \Omega_k$. Furthermore, we can also approximate $\phi_n (\omega \pm\Omega_k) \approx \phi_n(0)$ as before.
\begin{align}
    \cL_k [e^{i\sum_{\ell} \phi_{\ell}}\tilde\beta](\omega) 
    &=\frac{\kappa_k^2}{\Gamma_k^2} e^{i\sum_{\ell}\phi_{\ell} (0)} \Bigg[ \tilde\beta(\omega+\Omega_k)\mathcal{F}^{-1}_+[\tilde Q_k](0)^* \nonumber \\
    & \quad\quad  + \tilde\beta(\omega-\Omega_k)\mathcal{F}^{-1}_+[\tilde Q_k](0) \Bigg] ,
\end{align}
where $\mathcal{F}^{-1}_+[\tilde Q_k](0) = \int_0^\infty \diff \omega' \tilde Q_k(\omega')/\sqrt{2\pi}$ is the inverse Fourier transform of $\Theta(\omega)\tilde{Q}_k(\omega)$ at $t=0$ with $\Theta$ the Heaviside step function centered at $\omega=0$.
Plugged into the solution \eqref{eq:solution_betan_main}, we obtain a sum of all the sidebands from each cavity
\begin{align}
    \tilde\beta_N (\omega) &\approx e^{i\sum_{j=1}^N \phi_j (0)} \Big[\tilde\beta(\omega) \nonumber \\
    & \quad - i \sum_{k=1}^N \frac{g_k\kappa_k}{\Gamma_k^2}  e^{-i\phi_k(0)}\Big(\tilde\beta(\omega+\Omega_k)\mathcal{F}^{-1}_+[\tilde Q_k](0)^* \nonumber \\
    & \quad\quad  + \tilde\beta(\omega-\Omega_k)\mathcal{F}^{-1}_+[\tilde Q_k](0) \Big)\Big]. \label{eq:solution_betan_CW_cw_main}
\end{align}
In the simplest scenario, the mechanical frequencies are all the same, in which case the sideband amplitudes accumulate. The spectral profile of the sidebands is now given by the $\tilde\beta$-profile, as opposed to \eqref{eq:solution_betan_CW_main} where the profile was given by the mechanical signals, $\tilde Q_n$. \\
A summary of the approximations according to the different regimes is given in Table~\ref{tab:regimes}.

\begin{table*}[t]
\centering
\begin{tabular}{lccc}
\hline
 &
Stroboscopic &
CW &
CW (continuous signal) \\
\hline
Mechanical period vs. pulse duration &
$\Omega_n \tau \ll 1$ &
$\Omega_n \tau \gg 1$ &
$\Omega_n \tau \gg 1$ \\

Signal bandwidth vs. pulse duration &
-- &
$\Delta\Omega_n \tau \gg 1$&
$\Delta\Omega_n \tau\ll 1$\\
\hline
\end{tabular}
\caption{\label{tab:regimes} Restrictions for different considered limiting regimes. Additionally, $\kappa\tau\gg1$ is assumed in all of them.}
\end{table*}

\section{Coherent Enhancement of Metrological Bounds}\label{sec:metrology}

Based on the expressions for the output field of the $n$-th cavity and the cases discussed in the previous section, we now derive achievable lower bounds on  the uncertainty with
 which a signal encoded in the mechanical displacements $Q_n(t)$ can be estimated. We assume that all $Q_n$ are linearly proportional to a single parameter $\vartheta$ of the signal. A fundamental upper bound on the signal-to-noise ratio (SNR) for a given setup optimized over all possible measurements can be obtained via the Carm\'er-Rao bound as $\text{SNR}=\vartheta/\sqrt{\Delta\vartheta} \le \text{SNR}_{\rm CR} = \vartheta\sqrt{\mathcal{N}\mathcal{H}}$, where $\mathcal{N}$ is the number of independent repetitions of the measurement and $\mathcal{H}$ is the Quantum Fisher Information (QFI) \cite{pinel2013quantum} associated with the output field after the $N$-th cavity. Since we consider an initial coherent state of the light field and a quadratic Hamiltonian that  preserves Gaussianity, we can restrict our analysis generally to Gaussian states \cite{pinel2013quantum,vsafranek2018estimation}. The key advantage of Gaussian states lies in the fact that they are completely characterized by their first and second moments, collected in the displacement vector  $\boldsymbol{d} =\langle\hat{\boldsymbol{r}}\rangle^\intercal$ and the covariance matrix $\boldsymbol{\sigma}$ with components $\sigma_{ij}=\langle\{(\hat{r}_i-d_i),(\hat{r}_j^\dagger-d_j^*\}\rangle$, respectively. Here, $\hat{\boldsymbol{r}}=\{\ob^{(\rm out)}_N(\omega_1),\ob^{(\rm out)^\dagger}_N(\omega_1),...\}$ is the infinite dimensional vector of output creation and annihilation operators of the spectral modes which is obtained from the annihilation operators $\ob^{(\rm out)}_N(t)$ and the corresponding creation operators by Fourier transformation.

This structural property of Gaussian states allows the QFI to be written in a compact form, depending only on the complex covariance matrix and the displacement vector \cite{pinel2013quantum, ratzel2021decay}. 
In our model, the signal parameter is encoded only in the coherent field amplitudes and leaves the second moments unaffected. Therefore, we can restrict our analysis to the corresponding term in the QFI, 
\begin{equation}
    \mathcal{H}_{\bf d}=2(\partial_\vartheta\boldsymbol{d} )^\dagger \boldsymbol{\sigma}^{-1}(\partial_\vartheta\boldsymbol{d} ).\label{eq:QFIdef}
\end{equation}
The associated Cram\'er-Rao bound can be saturated by spectrally (or temporally) resolved homodyne detection of the output field.
For non-Gaussian field states and homodyne measurements, the CFI of the measured quadrature distribution is lower bounded by the expression~\eqref{eq:QFIdef}.

In the following, we assume that the covariance matrix is that of the free field, $\boldsymbol{\sigma} = \mathbb{I}$.
The QFI for a continuum of spectral modes reduces to
\begin{align}
        \mathcal{H}_{\bf d} 
        & = 4\!\!\int\text{d}\omega|\partial_\vartheta\langle \hat{b}^{(\rm out)}_N(\omega)\rangle|^2  = 4\eta^{N-1} \!\! \int\text{d}\omega|\partial_\vartheta\tilde\beta_N(\omega)|^2 .
\end{align}
A complete treatment would include perturbations of the covariance matrix due to thermal fluctuations of the mechanical resonator and the back-action of the readout light on the mechanics. In the weak coupling regime ($g_n{\ll} \kappa_n$), back-action can be safely neglected as the effect on the light field is of second order in $g_n$. We do not assume that the mechanical elements are cooled to the quantum ground state but consider the opposite regime where the environmental temperature is much larger than $\hbar\Omega_n/k_{\rm B}$ for all $n$, where $k_{\rm B}$ is Boltzmann's constant. Therefore, thermal fluctuations dominate over quantum fluctuations. An analysis of thermal contributions to the covariance matrix is given in Appendix~\ref{app:corrcovmatrix} using the stroboscopic regime as an example. In that case, we find that $\boldsymbol{\sigma} = \mathbb{I}$ is a good approximation if the mechanical element is cooled to temperatures of several Kelvin for typical optomechanical parameters. 

\subsection{SNR for three limiting cases}\label{subsec:SNRresults}

We immediately obtain bounds on the SNR after $N$ optomechanical cavities for the different regimes discussed above. We give the expressions for single-shot experiments ($\mathcal{N}=1$) as independent repetitions only give rise to a factor $\sqrt{\mathcal{N}}$. For the stroboscopic regime, we find
\begin{equation}
    \begin{aligned}
        \text{SNR}_{\rm st}(N) &\le 2\eta^{\frac{N-1}{2}} \vartheta\left(\int\text{d}\omega|\partial_\vartheta\tilde\beta_N(\omega)|^2\right)^{1/2}\\
        &=   4 \eta^{\frac{N-1}{2}} \Bigg(\int\text{d}\omega\Bigg|\sum_{j=1}^N \frac{g_j}{\kappa_j} Q_j(0)\times  \\ 
        &   \quad\quad\quad \times  \left( 1 - \cos\phi_j(\omega)\right)\tilde\beta(\omega)\Bigg|^2\Bigg)^{1/2}.
    \end{aligned}
\end{equation}
In addition, if the pulse spectrum and the cavity detuning are chosen such that the pulse fits into each of the cavity modes, that is, $\tau^{-1},\Delta_n \ll \kappa_n$, we can evaluate the phases $\phi_j$ at $\omega=0$ (the peak of $\tilde\beta$), and
\begin{equation}
    \begin{aligned}
        \text{SNR}_{\rm st}(N) &\le 4 \eta^{\frac{N-1}{2}} N_{\rm in}^{1/2}\left|\sum_{j=1}^N \frac{g_j}{\kappa_j} Q_j(0) \left( 1 - \cos\phi_j(0)\right)\right|,
    \end{aligned}
\end{equation}
where $N_{\rm in}=\sqrt{\pi}\tau\bar{\beta}^2$ is the number of input photons. We see that the SNR is additive in the case of vanishing noise, $\eta\rightarrow 1$.

In the continuous wave regime with finite signal duration, we obtain
\begin{equation}
    \begin{aligned}
        \text{SNR}_{\rm cw}(N) &\le 2 \eta^{\frac{N-1}{2}} N_{\rm in}^{1/2} \frac{1}{\sqrt{\tau\sqrt{\pi}}} \times \\
        & \quad \times \left(\int\text{d}\omega\left| \sum_{k=1}^N
        \frac{g_k\kappa_k}{\Gamma_k^2}e^{-i\phi_k(0)} \tilde Q_k (\omega) \right|^2\right)^{1/2} 
    \end{aligned}
\end{equation}
If the spectral sidebands produced by each optomechanical cavity are assumed to be equivalent up to their amplitude, that is, $\tilde{Q}_k(\omega)=\tilde{Q}_{k,0}\tilde{S}(\omega)$, the integration over $|\tilde{S}(\omega)|^2$ can be performed independently from the sum over cavities. This results in
\begin{equation}
    \begin{aligned}
        \text{SNR}(N) &\le 2 \eta^{\frac{N-1}{2}} N_{\rm in}^{1/2}  \mathcal{M}_{\rm sb}\left|\sum_{k=1}^N \frac{g_k\kappa_k}{\Gamma_k^2}e^{-i\phi_k(0)} \tilde Q_{k,0}\right|
    \end{aligned}
\end{equation}
where $\mathcal{M}_{\rm sb}=(\int\text{d}\omega|\tilde{S}(\omega)|^2)^{1/2}/\sqrt{\tau\sqrt{\pi}}$.
For the case of a continuous signal, we use Eq.~\eqref{eq:solution_betan_CW_cw_main} and take into account that the two sidebands do not overlap (since $\Omega_k\tau\gg 1$) to obtain
\begin{equation}
    \begin{aligned}
        \text{SNR}(N) &\le 2 \eta^{\frac{N-1}{2}} N_{\rm in}^{1/2} \times \\
        & \quad \times \Bigg(\left| \sum_{k=1}^N \frac{g_k\kappa_k}{\Gamma_k^2}e^{-i\phi_k(0)} \mathcal{F}^{-1}_+[\tilde Q_k](0)^* \right|^2 \\
        & \quad + \left| \sum_{k=1}^N \frac{g_k\kappa_k}{\Gamma_k^2}e^{-i\phi_k(0)} \mathcal{F}^{-1}_+[\tilde Q_k](0) \right|^2       
        \Bigg)^{1/2} 
    \end{aligned}
\end{equation}
This concludes the limiting cases that we treat in this article. In the following we restrict our considerations to equal cavities which leads to general results that apply to all limiting cases.

\subsection{Equal cavities}\label{subsec:equalcav}

\begin{figure}
    \centering
    \includegraphics[width=0.9\linewidth]{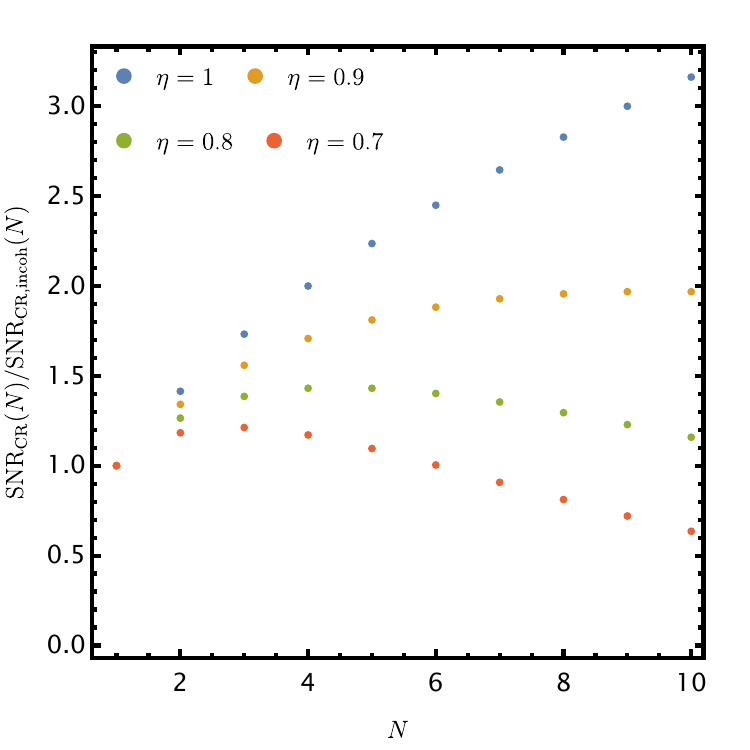}
    \caption{Plot of the ratio of the upper bounds for the coherent and incoherent signal to noise ratios $\text{SNR}_{\rm CR}(N)$ and $\text{SNR}_{\rm CR, incoh}(N)$. In the ideal case of no losses (blue dots) the ratio increases indefinitely proportionally to $\sqrt{N}$. As losses are introduced (yellow dots $\eta=0.9$, green dots $\eta=0.8$, red dots $\eta=0.7$) there is an interplay between the coherent accumulation of the signal and decoherence which leads to an optimal number of systems $N_{\rm opt}$ above which the advantage is lost, and the SNR decreases. Higher values of losses correspond to lower SNR-ratios and smaller $N_{opt}$. }
    \label{SNRratio-plot}
\end{figure}

If we assume that all  sensors provide the same contribution to the SNR, we obtain the upper bound for the SNR after $N$ sensors as
\begin{equation}\label{eq:SNRgain_idCav}
    \text{SNR}_{\rm CR}(N)= N\eta^{\frac{N-1}{2}} \text{SNR}_{\rm CR}(1)
\end{equation} 
for all three cases, where $\text{SNR}_{\rm CR}(1)$ is the bound of the SNR for a single optomechanical cavity. 
To evaluate the advantage of the cascaded sensing scheme in comparison to incoherent combination of the measurement outcomes of $N$ optomechanical sensors into one measurement statistics, we have to compare the SNR scaling above with the scaling of independent measurements $\text{SNR}_{\rm CR, incoh}(N)= \sqrt{N}\text{SNR}_{\rm CR}(1)$. In Fig.~\ref{SNRratio-plot}, the quotient $\text{SNR}_{\rm CR}(N)/\text{SNR}_{\rm CR, incoh}(N)=\sqrt{N}\eta^{(N-1)/2}$ is plotted as function of
$N$  for different values of $\eta$. While the SNR increases indefinitely with $N$ for attenuation parameter $\eta=1$, for each $\eta\neq 1$, there exists an $N_{\rm opt}$ at which $\text{SNR}_{\rm CR}(N)/\text{SNR}_{\rm CR, incoh}(N)$ reaches a maximum. Without any further noise sources and assuming that measurements are implemented that actually saturate the bound (e.g. homodyning ones), $N_{\rm opt}$ is the number of optomechanical cavities to optimally exploit the cascaded sensing scheme. It is easy to show analytically that $N_{\rm opt}$ is one of the two integers that are closest to $-\log(\eta)^{-1}$. For $\eta$ very close to 1 (roughly for values larger $0.7$), we find that $N_{\rm opt}$ is the integer part of $\eta/(1-\eta)$ and $\text{SNR}_{\rm CR}(N_{\rm opt})/\text{SNR}_{\rm CR, incoh}(N_{\rm opt})\approx 1/\sqrt{e(1-\eta)}$, where $e$ is Euler's number.

\begin{figure}
    \centering
    \includegraphics[width=1\linewidth]{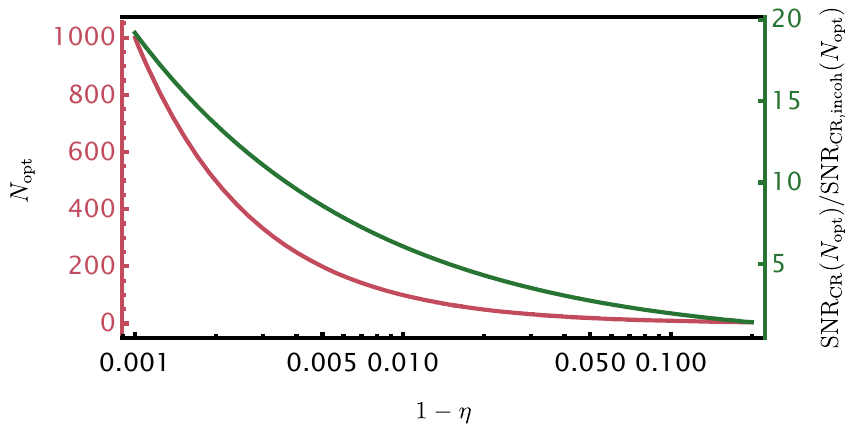}
    \vspace*{-0.5cm}
    \caption{Plots of the optimal number of sensors $N_{\rm opt}$ and the maximal SNR ratio $\text{SNR}_{\rm CR}(N_{\rm opt})/\text{SNR}_{\rm CR, incoh}(N_{\rm opt})$ as a function of the loss from the light field between the sensors $1-\eta$ from $\eta=0.999$ to $\eta=0.8$.
    }  
    \label{fig:eta_variation}
\end{figure}

In Fig.~\ref{fig:eta_variation}, we show plots of the optimal number of sensors $N_{\rm opt}$ and the corresponding enhancement ratio  $\text{SNR}_{\rm CR}(N_{\rm opt})/\text{SNR}_{\rm CR, incoh}(N_{\rm opt})$ for values of $\eta$ between $0.8$ and $0.999$.   

We like to note that, in practice, remaining differences between cavity frequencies and small errors in the adjustments of the optical delays between the sensors are unavoidable. While the latter approximately correspond to constant phase shifts (under the assumption that the product of signal bandwidth and delay timing error is much smaller than one), mechanical frequency differences lead to a dephasing that increases with time in the case of continuous measurements and have to be adjusted regularly. The resulting phase shifts can be modeled as stochastic fluctuations with a vanishing mean. As can be seen from our results in Sec.\ref{subsec:SNRresults}, phase differences between the amplitudes from different cavities enter the SNR essentially as a factor $|\sum_n e^{i\delta \phi_n}|$ (under the assumption of otherwise identical cavities). For phase fluctuations much smaller than 1, this expression can be expanded to second order and averaged. For fluctuations of the mechanical phase, we can assume identical variance which will not affect the $N$-scaling to leading order. The phase fluctuations resulting from optical delay fluctuations $\delta T_n$ will be approximately of the form $\delta\phi_n=\Omega\sum_{n=1}^{N-1} \delta T_n$, where $\Omega$ is the average mechanical frequency. If we assume equivalent variance $\sigma_T^2$  of the $\delta T_n$, the $N$-scaling is ensured to be approximately preserved for $N\Omega^2\sigma_T^2\ll 1$.

\section{Potential Applications}\label{sec:applications}

In this section, we discuss three sensing applications which could profit from the cascaded sensing scheme presented in this article.

\subsection{Dark Matter}\label{subsec:DM}

Despite the abundance of astrophysical and cosmological observations in favor of the existence of Dark Matter (DM) particles, the fundamental nature of these constituents has not been identified. DM is expected to interact very weakly with standard matter (SM) and perhaps only through gravity.
A crucial characteristic such as the mass has not been established yet, and the range of possible values is very broad, from $10^{-58}~\mathrm{kg}$ up to several of solar masses. Observations indicate that the local dark matter (DM) mass density is $\rho_\chi\simeq 5.3 \times10^{-22}~\mathrm{kg/m^3}$
~\cite{carney2021mechanical}.
DM candidates are usually classified according to their masses $m_\chi$ into two main categories and their interaction with standard matter can be described in different ways: heavy DM or particle-like candidates defined by $m_\chi>10^{-36}~\mathrm{kg}$
are expected to behave as a dilute gas of single particles; ultralight wave-like candidates, with $10^{-58}~\mathrm{kg}<m_\chi<0.1\times 10^{-36}~\mathrm{kg}$
are described as a persistent wave-like field. Different experimental setups and measurement apparata are required to investigate different parts of the parameter space. However, for all the models interacting with SM only through mass, massive mechanical sensors are required, and optomechanics offer potential platforms to search a wide range of candidates.
Assuming that the DM is in equilibrium with the galaxy halo, it is expected to move at a typical speed of $v_\chi\sim 10^5~\mathrm{m/s}$, from the viral theorem, and then having a de Broglie wavelength of $\lambda_\chi=2\pi\hbar/(m_\chi v_\chi)$~\cite{carney2021ultralight}. We focus on ``ultra light" DM, where candidates possess low mass ($m_\chi<0.1 \times 10^{-36}~\mathrm{kg}$) and extremely large mode occupation numbers. This means that we can treat these candidates as a bosonic field that oscillates in time with wavelength $\lambda_\chi>1~\mathrm{mm}$  and natural field frequency $\omega_\chi = m_\chi c^2/\hbar \lesssim 2\pi \times 10^{13}~\mathrm{s^{-1}}$.
The frequency spread $\delta\omega_\chi\approx m_\chi v^2_\chi/(2\hbar)\approx 2\pi \times10^{6}\mathrm{s}^{-1}$ leads to a coherence time $T_{\rm coh}\approx 2\pi/\delta \omega_\chi\approx 2\pi\times10^7/\omega_\chi$ ~\cite{carney2021mechanical}. These models therefore give rise to extremely weak, coherent and persistent field oscillations that can potentially be detected, and the core problem reduces to the detection of a weak signal which makes it an extremely suitable application for the cascaded optomechanical scheme. Assuming that DM consists of a single ultralight field, the observable signature is a temporally oscillating force that acts on the detector, with strength proportional to the mass. This can be parametrised as
\begin{equation}
    F_\chi=g_\chi \rho_\chi f(t),
\end{equation}
where $g_\chi$ is the unknown DM coupling that depends on the DM model assumed, $\rho_\chi$ is the local DM density, and $f(t)$ contains the time-dependence of the signal. In the steady state at resonance, the resulting amplitude of a mechanical sensor with frequency $\Omega$ and damping rate $\Omega\zeta$ is $q_{\chi,{\rm s}}= F_\chi/(2m\Omega^2\zeta)$. Hence, a search for DM could be done by trying to estimate the parameter $\vartheta_\chi= g_\chi \rho_\chi$ by optomechanical read-out of the mechanical amplitude. This approach is promising, as it is supported by several studies on optomechanical methods for DM detection \cite{carney2021mechanical,qin2025mechanical, monteiro2020search, manley2020searching, hamaide2025searching}. Since some sensors are small systems, it has also been already proposed to use arrays of many of these sensors \cite{carney2021mechanical}. Due to the small size of the systems and the possibility to combine them with integrated optics, losses between sensors may be significantly restricted. In that case, our framework of cascaded optomechanical sensing is applicable and the coherent enhancement of sensitivity provides a factor $\sqrt{N}$ in the most ideal case.

\subsection{Gravitational waves}\label{subsec:GW}

Since the first detection of gravitational waves by the LIGO/Virgo collaboration~\cite{abbott2016observation}, many more transient signals from binary mergers have been observed and gravitational astrophysics is expanding quickly with new detectors going online and being proposed. Also continuous gravitational wave signals from large binaries in the low frequency regime are very likely to be detected soon with LISA in space~\cite{wyithe2003low}. However, signals from binary mergers only represent a subset of possible signals. Other types include non-transient gravitational wave signals in the kHz-band and above that can be continuous, for example, emitted by fast rotating neutron starts~\cite{lasky2015gravitational,Abbott2019allsky,dergachev2021results}, or a stochastic background, for example, emitted during the rapid evaporation of primordial black holes~\cite{christensen2018stochastic}. 

Besides the general possibility to employ interferometric gravitational wave antennas like LIGO and Virgo for the detection of non-transient gravitational wave signals \cite{christensen2018stochastic,Abbott2019allsky}, there are several other proposals to employ highly controlled mechanical sensors for the detection of continuous gravitational wave signals at frequencies in the kHz range and above 
\cite{arvanitaki2013detecting,Goryachev2014grav,Singh:2016xwa,Goryachev2021rare,page2021gravitational,tobar2025detecting}. Since these sensors are usually much smaller than the wavelength of the gravitational waves they are proposed to detect, in essence, a gravitational wave of frequency $\omega$ and strain $h$ acts as a time dependent tidal force with gradient \cite{maggiore2008gravitational,ratzel2019testing}
\begin{equation}
    |\nabla F_h| \propto \omega^2 h/2\,.
\end{equation}
In many cases, the proposed readout mechanism for the mechanical sensors is via the electromagnetic field and sensors can be bundled into a larger volume to form a more sensitive detector network (as has been demonstrated for torsion bar antennas \cite{Shoda2014search} and interferometers \cite{patra2025broadband}). Therefore, gravitational wave detection of continuous high-frequency signals is another potential application for cascaded optomechanical metrology.

\subsection{Gravitational field of ultra-relativistic matter}\label{subsec:LHC}

A third potential application of the cascaded optomechanical sensing scheme is the detection of the gravitational field of ultra-relativistic matter at the Large Hadron Collider (LHC)~\cite{spengler2022perspectives}. The gravitational attraction of the circulating proton bunches is dominated by their kinetic energy which is a factor $\sim 7000$ larger than the energy associated with their rest mass. While the gravitational effect of kinetic energy is evidenced in many astrophysical observations, this prediction of general relativity has not been experimentally tested. Tests at the LHC would enable full control over the generation of ultra-relativistic sources of gravitational fields, allowing for precise and repeatable measurements~\cite{braun2025path}.
Besides the exciting possibility of the first direct evidence for the gravitational attraction of kinetic energy, gravity measurements at particle accelerators could also be used to test alternative gravity theories in yet unexplored regimes \cite{pfeifer2025gravitational}.
The LHC consists of a $27$ km ring in which $2800$ bunches of protons circulate at speeds approaching $c$, with each bunch carrying a total energy of $10^5$ J distributed among approximately $1.15\times 10^{11}$ protons. Each bunch is approximately $30$ cm long and lasts approximately $1$ ns. To profit from phase locking techniques, a sensor positioned near the ring must resonate at frequencies matching the bunch passage rate: the full beam provides a rate of $31.2$ MHz, while a single bunch orbiting the ring passes with a frequency of $11$ kHz. Lower frequencies can be achieved through periodic modulation of the beam position, where the LHC beam is active for half the sensor's oscillation period. 
One can estimate the effective acceleration amplitude driving the sensor as~\cite{spengler2022perspectives}
\begin{equation}
    a_{0,\rm LHC} \sim \dfrac{4 GP}{c^2 d},
\end{equation}
where $G$ is the gravitational constant, $P$ the average power of the particle beam ($3.8 \times 10^{12}$ W in the case of the LHC) and $d$ the distance between the center of the mechanical resonator and the beam axis. This leads to the corresponding steady state oscillation amplitude $q_{\chi,{\rm s}}= a_{\rm LHC}/(2\Omega^2\zeta)$ of the mechanical resonator.

Since the proton bunches at the LHC move in a 27 km long ring, in principle, many acceleration sensors can be placed around the beam line and their signals can be correlated. Therefore, significant enhancement could be provided by the cascaded sensing scheme if losses between sensors can be limited.

\section{Conclusions}\label{sec:conclusions}

We have shown that coherently coupling $N$ optomechanical sensors via their output light field in a cascaded way can improve the total signal-to-noise ratio (SNR) significantly. In the ideal case, that is, without loss from the output light field between the cavities, this version of coherent averaging leads to an increase of the SNR proportional to $N$ while incoherent averaging leads to the usual scaling with $\sqrt{N}$. If losses are included, represented by an attenuation factor $\eta$, the advantage of the cascaded readout over the incoherent readout scales as $\eta^{(N-1)/2}\sqrt{N}$ which implies an optimal value for $N$ beyond which the advantage starts to decrease. For $\eta$ close to one, the enhancement factor is approximately $1/\sqrt{e(1-\eta)}$, where $e$ is Eulers number, which can mean a significant improvement. Note that coherent averaging can be combined with quantum resources such as squeezed input light to increase sensitivity further \cite{brady2023entanglement,xia2023entanglement}.

We have discussed that the cascaded sensing scheme can be employed in fundamental research with the examples of dark matter detection, gravitational wave astronomy and measurements of the gravitational field of ultra-relativistic matter. We expect that further proposals for applications may be found in the future.

In this work we modeled the mechanical oscillator as a moving end mirror of a cavity without considering back-action from the readout light. While this is justified for the case of macroscopic mirrors and small back-action (as we show in the appendix), it may be worthwhile to extend our analysis to a full quantum mechanical model treating the mirror position as another dynamical degree of freedom. This is left for future work.

\subsection*{Acknowledgements}

We thank Germain Tobar, Navdeep Arya, Magdalena Zych, Teodor Strömberg  and Onur Hosten for useful discussions. 
MMM acknowledges support from the Walter Benjamin Programme (project number 510053905) and from the Austrian Science Fund (FWF) [10.55776/COE1]. DB acknowledges funding by the EU EIC Pathfinder project QuCoM (101046973). D.R. acknowledges funding by the Federal Ministry of Education and Research of Germany in the project “Open6GHub” (grant number: 16KISK016),  financial support from EPSRC (Engineering \& Physical Sciences Research Council, United Kingdom) Grant Number EP/X009467/1, and support by the Deutsche Forschungsgemeinschaft (DFG, German Research Foundation) under Germany’s Excellence Strategy – EXC-2123 QuantumFrontiers – 390837967.

\newpage
\onecolumngrid
\appendix

\section*{Appendix}

\section{General recursion formula}
\label{app:recursion}

We begin from the Hamiltonian $\oH_n$ describing each one-sided cavity, written in a frame rotating at the (central) probe frequency $\omega_L$. Equivalently, one may transform into the frame co-rotating with the input light by applying the unitary $\oU_n(t)=\exp \{i\omega_L [\oa^\dagger_n\oa_n+\int d\omega \ob_n^\dagger(\omega)\ob_n(\omega)]t\}$.
In this frame, the bare cavity resonance $\omega_n$ appears through the detuning $\Delta_n=\omega_n-\omega_L$ and the linewidth $\kappa_n$. The mechanical signal is treated classically and encoded in the dimensionless quadrature $q_n(t)$, oscillating at a mechanical frequency $\Omega_n$. We work in the bad-cavity limit $\Omega_n\ll\kappa_n$, and assume weak optomechanical coupling strengths $g_n$.

The equations of motion for the co-rotating field operators read
\begin{align}
\partial_t \oa_n (t) &= -i\Delta_n \oa_n (t) + i g_n q_n(t) \oa_n(t) - \sqrt{\frac{\kappa_n}{2\pi}}\int \diff \omega  \, \ob_n (\omega,t), \nonumber \\
\partial_t \ob_n (\omega,t) &= \sqrt{\frac{\kappa_n}{2\pi}} \oa_n (t) - i(\omega-\omega_L)\ob_n (\omega,t) ,
\end{align}
where $t_0\to -\infty$ denotes the asymptotic initial time at which the cavity and the external field are uncoupled, so that $[\oa_n(t_0),\ob_n^\da (\omega,t_0)]=0$. The continuum of operators outside the cavity satisfies the standard commutation relations $[\ob_n(\omega,t),\ob_n^\da (\omega',t)]=\delta(\omega-\omega')$ at all times. Under the Markov approximation of a flat coupling density, one can integrate the second equation from $t_0$ to $t$ and substitute the result into the first equation, yielding
\begin{equation}\label{eq:an_dt}
    \partial_t \oa_n (t) = \left[-i\Delta_n - \frac{\kappa_n}{2} +ig_nq_n(t) \right] \oa_n(t) - \sqrt{\kappa_n} \ob_n^{\rm (in)} (t), \quad \ob_n^{\rm (in)}(t) = \int \frac{\diff \omega}{\sqrt{2\pi}}e^{-i(\omega-\omega_L)(t-t_0)} \ob_n (\omega,t_0).
\end{equation}

Alternatively, the cavity dynamics may be written in terms of the output field $\ob_n^{\rm (out)}(t) = (2\pi)^{-1/2}\int \diff\omega\, e^{-i(\omega-\omega_L)(t-t_1)}\ob_n(\omega;t_1)$ with $t_1\to\infty$, leading to the usual input–output relation~\cite{gardiner1985input}
\begin{equation}\label{eq:IO_bn}
    \ob_n^{\rm (out)} (t) = \ob_n^{\rm (in)} (t) + \sqrt{\kappa_n} \oa_n (t).
\end{equation}
The corresponding Heisenberg-picture operators in the laboratory frame are $\oa_n^{\rm (H)}(t)=\oa_n(t)e^{-i\omega_L t}$ and $\ob_n^{\rm (H)}(\omega,t)=\ob_n(\omega,t)e^{-i\omega_L t}$. This representation is useful when evaluating expectation values with respect to the initial state of the incoming field before the first cavity,
\begin{equation}\label{eq:psi_in}
    |\psi (t_0)\ra = \prod_\omega \oD_1 \left[ \beta_0 \tau e^{-i\omega t_0 - \tau^2(\omega-\omega_L)^2/2} \right] |\vac\ra,
\end{equation}
which corresponds to a Gaussian pulse of duration $\tau$, centered at $\omega_L$, and arriving at the first cavity at $t=0$. One then finds
\begin{equation}
    \la \ob_1^{\rm (in,H)} (t) \ra = \int\frac{\diff\omega}{\sqrt{2\pi}} e^{-i\omega(t-t_0)}\la \psi (t_0)| \ob_1^{\rm (H)}(\omega,t_0)|\psi(t_0)\ra = \beta_0 e^{-t^2/2\tau^2-i\omega_L t}, \quad \text{so that}\ \ \beta(t):=\la \ob_1^{\rm (in)} (t) \ra = \beta_0 e^{-t^2/2\tau^2}.
\end{equation}
The continuous-wave limit is recovered by taking $\tau\to\infty$.

In the cascaded configuration, the asymptotic output of cavity $(n-1)$ serves as the input to cavity $n$ after free propagation over a delay time $T\gg 1/\kappa$. In the laboratory frame this implies $\ob_n^{\rm (in,H)}(t)=\ob_{n-1}^{\rm (out,H)}(t-T)$, and in the co-rotating frame,
\begin{equation}\label{eq:OI_n_np1}
 \ob_n^{\rm (in)} (t) = \ob_{n-1}^{\rm (out)} (t-T)e^{i\omega_L T}  , \quad n>1.
\end{equation}
By repeatedly applying Eqs.~\eqref{eq:IO_bn} and \eqref{eq:OI_n_np1}, the output field of the $n$th cavity can be written in terms of the original input field and the accumulated response of all preceding cavities,
\begin{equation}
    \ob_n^{\rm (out)} (t) = \ob_{n-1}^{\rm (out)} (t-T)e^{i\omega_L T} + \sqrt{\kappa_n}\oa_n (t) = \ldots=\ob_1^{\rm (in)}(t-(n-1)T)e^{i(n-1)\omega_L T} + \underbrace{\sum_{j=0}^{n-1} \sqrt{\kappa_{n-j}} \oa_{n-j} (t-jT) e^{ij\omega_L T}}_{=\oB_n(t-(n-1)T)e^{i(n-1)\omega_L T}}.
\end{equation}
The cumulative response up to cavity $n$ is therefore defined as
\begin{equation}\label{eq:Bn_def}
    \oB_n (t) = \sum_{j=0}^{n-1}\sqrt{\kappa_{n-j}} \oa_{n-j} (t+(n-j-1)T) e^{-i(n-j-1)\omega_L T} = \sum_{k=1}^{n}\sqrt{\kappa_k} \oa_k (t+(k-1)T) e^{-i(k-1)\omega_L T} .
\end{equation}

Ideally, the mechanical modulation should be synchronized with the arrival of the pulse at the $n$th cavity $(n-1)T$. We therefore write $q_n(t)=Q_n(t-(n-1)T)$, with $Q_n(t)$ centered around $t=0$. Introducing $\Gamma_n=\kappa_n/2+i\Delta_n$, we integrate the Langevin equation \eqref{eq:an_dt} to obtain
\begin{align}
    \oa_n (t) =&\ \oa_n(t_0) \exp \left[-\Gamma_n(t-t_0) +ig_n \int_{t_0}^t \diff t'\, Q_n (t'-(n-1)T) \right]  \\
    & - \sqrt{\kappa_n}\int_{t_0}^t \diff t'\, \ob_n^{\rm (in)} (t') \exp \left[-\Gamma_n(t-t') +ig_n \int_{t'}^t \diff t''\, Q_n (t''-(n-1)T) \right] \nonumber \\
    \approx&\ \oa_n(t_0) e^{-G_n (t-(n-1)T)(t-t_0)} - \sqrt{\kappa_n} \int_{0}^{t-t_0} \diff t'\, \ob_n^{\rm (in)} (t-t')e^{-G_n(t-(n-1)T)t'}, \quad \text{with }\ G_n(t) := \Gamma_n - ig_n Q_n (t). \nonumber
\end{align}
In the approximation, we have used the bad-cavity condition: $Q_n (t') \approx Q_n(t)$ as long as $t'$ lies within the response time of the cavity, $|t-t'| \lesssim 1/\kappa_n$.

Next, we shift the time argument by the pulse arrival time and rewrite the result in terms of the first input field and the cumulative response of the earlier cavities,
\begin{align}
   \oa_n (t+(n-1)T) &=  \oa_n(t_0) e^{-G_n (t)(t+(n-1)T-t_0)} - \sqrt{\kappa_n} \int_{0}^{t+(n-1)T-t_0} \diff t'\, \ob_{n-1}^{\rm (out)} (t + (n-2)T-t')e^{-G_n(t)t'+i\omega_L T} \nonumber \\
   &= \oa_n(t_0) e^{-G_n (t)(t+(n-1)T-t_0)} - \sqrt{\kappa_n} \int_{0}^{t+(n-1)T-t_0} \diff t' \, e^{-G_n(t)t'+i(n-1)\omega_L T} \left[ \ob_1^{\rm (in)} (t-t') + \oB_{n-1}(t-t')\right] \nonumber \\
   &\xrightarrow{t_0 \to-\infty} - \sqrt{\kappa_n}e^{i(n-1)\omega_L T} \int_{0}^{\infty} \diff t' \, e^{-G_n(t)t'} \left[ \ob_1^{\rm (in)} (t-t') + \oB_{n-1}(t-t')\right] \qquad \text{for any finite }t.
\end{align}
This limit is justified as long as the cavity field is evaluated at finite times, well after the asymptotic past $t_0$.

To derive a recursion relation for $\oB_n$, we sum according to Eq.~\eqref{eq:Bn_def},
\begin{align}
    \oB_n (t) &= - \sum_{k=1}^n \kappa_k \int_{0}^{\infty} \diff t' \, e^{-G_k(t)t'} \left[ \ob_1^{\rm (in)} (t-t') + \oB_{k-1}(t-t')\right] \nonumber \\
    &= -\kappa_n \int_{0}^{\infty} \diff t' \, e^{-G_n(t)t'} \left[ \ob_1^{\rm (in)} (t-t') + \oB_{n-1}(t-t')\right]\ \underbrace{- \sum_{k=1}^{n-1} \kappa_k (\ldots)}_{=\oB_{n-1}(t)} \nonumber \\
    &= \oB_{n-1} (t) -\kappa_n \int_{0}^{\infty} \diff t' \, e^{-G_n(t)t'} \left[ \ob_1^{\rm (in)} (t-t') + \oB_{n-1}(t-t')\right].
    \label{eq:Bn_recursion}
\end{align}
With $\oB_0\equiv 0$, this recursion holds for any $n\ge 1$. Because of the explicit time dependence in the exponential, the integral does not reduce to a simple convolution, and in general no closed-form analytic solution is expected.

Our main interest lies in the expectation value of the cumulative response of a large number $N\gg 1$ of cavities, $\la \oB_N(t)\ra$, as it is the coherent output field amplitude by which we measure the mechanical signal. It is convenient to introduce the time-shifted output amplitude $\beta_n(t)=\beta(t)+\la \oB_n(t)\ra$, for which the recursion becomes
\begin{equation}\label{eq:betan_recursion}
    \la \ob_n^{\rm (out)} (t)\ra = \beta_n (t-(n-1)T)e^{i(n-1)\omega_L T} \qquad \Rightarrow \qquad \beta_n(t) = \beta_{n-1} (t) - \kappa_n \int_{0}^{\infty} \diff t' \, e^{-G_n(t)t'} \beta_{n-1} (t-t'),
\end{equation}
with $\beta_0\equiv\beta$. The measured output in the laboratory frame and its Fourier components are given by
\begin{equation}
   \la \ob_n^{\rm (out,H)} (t)\ra = \beta_n (t-(n-1)T)e^{i\omega_L[(n-1)T-t]}, \qquad  \la \ob_n^{\rm (out,H)} (\omega)\ra = \int\frac{\diff t}{\sqrt{2\pi}}e^{i\omega t}\la \ob_n^{\rm (out,H)} (t)\ra = \tilde\beta_n (\omega-\omega_L)e^{i(n-1)\omega T} .
\end{equation}
Here $\tilde\beta_n(\omega)=\int \diff t\, e^{i\omega t}\beta_n(t)/\sqrt{2\pi}$. Using the inverse Fourier transform, Eq.~\eqref{eq:betan_recursion} can be expressed in frequency space as
\begin{align}
\tilde\beta_n(\omega) &=  \tilde\beta_{n-1}(\omega) - \kappa_n \int\frac{\diff t}{\sqrt{2\pi}}e^{i\omega t} \int_0^{\infty}\diff t'\, e^{-G_n(t)t'} \int \frac{\diff \omega'}{\sqrt{2\pi}} e^{-i\omega'(t-t')} \tilde\beta_{n-1}(\omega') \nonumber \\
&=  \tilde\beta_{n-1}(\omega) - \frac{\kappa_n}{2\pi} \int\diff\omega' \tilde\beta_{n-1}(\omega') \underbrace{\int\diff t \, \frac{e^{i (\omega-\omega')t}}{G_n (t)-i\omega'}}_{=: F_n (\omega',\omega-\omega')} .
\end{align} 
Since realistic optomechanical couplings satisfy $g_n\ll\kappa_n$, and we are interested in weak mechanical signals, it is natural to expand to lowest order in $|g_n Q_n(t)|/\kappa_n$,
\begin{equation}
    F_n(\omega',\Omega) \approx \frac{2\pi \delta(\Omega)}{\Gamma_n - i\omega'} + \frac{\sqrt{2\pi}ig_n \tilde Q_n(\Omega)}{(\Gamma_n - i\omega')^2}, \qquad \text{with}\quad \tilde Q_n(\Omega) = \int\frac{\diff t}{\sqrt{2\pi}} Q_n(t) e^{i\Omega t} = \tilde q_n (\Omega) e^{-i\Omega (n-1)T}.
\end{equation}
This first-order expansion is sufficient to evaluate the quantum Fisher information around vanishing mechanical signal amplitude.

The mechanical excitation oscillates at a frequency $\Omega_n\ll\kappa_n$, but it need not be strictly harmonic. One may instead consider a mechanical pulse with spectral profile $\tilde Q_n(\Omega)$ centered at $\Omega_n$ and width $\Delta\Omega_n<\Omega_n$.
Introducing the cavity phase response $e^{i\phi_n(\omega)}=-(\Gamma_n-i\omega)^*/(\Gamma_n-i\omega)$, the recursion becomes
\begin{equation}
    \tilde\beta_n (\omega) = e^{i\phi_n(\omega)}\tilde\beta_{n-1} (\omega) - i \frac{\kappa_n g_n}{\sqrt{2\pi}} \int\diff \omega' \frac{\tilde\beta_{n-1}(\omega')}{(\Gamma_n - i\omega')^2} \tilde Q_n (\omega-\omega') + \Order \{\eps_n^2 \}, \qquad \tilde\beta_0(\omega)=\tilde\beta(\omega).
\end{equation}
The integral term is of order $\eps_n:=g_n/\kappa_n$ and represents a convolution in frequency, with a width determined by the broader of the two integrands. Iterating this relation while consistently retaining only first-order terms in $\eps_n$ yields the total response.

For compactness, we define the linear operator $\cL_n$ acting on a frequency-domain function as
\begin{equation}\label{eq:Ln}
    \cL_n [\tilde\alpha](\omega) := \int\diff\omega' \frac{\kappa_n^2 \tilde\alpha(\omega')}{(\Gamma_n-i\omega')^2} \frac{\tilde Q_n(\omega-\omega') }{\sqrt{2\pi}} = \int\diff\omega' \left[1 - e^{i\phi_n(\omega')} \right]^2 \tilde\alpha (\omega') \frac{\tilde Q_n(\omega-\omega') }{\sqrt{2\pi}}.
\end{equation}
In this notation, the recursion reads
\begin{equation}
    \tilde\beta_n = e^{i\phi_n} \tilde\beta_{n-1} - i\eps_n \cL_n [\tilde\beta_{n-1}], \qquad \tilde\beta_0 = \tilde\beta,
\end{equation}
where $\tilde\beta_k$ and $\phi_k$ are functions of $\omega$. The consistent first-order solution is
\begin{equation}\label{eq:solution_betan}
    \tilde\beta_n = \exp \left\{i\sum_{j=1}^n \phi_j \right\} \tilde\beta - i\sum_{k=1}^n \eps_k \exp \left\{i\sum_{j=k+1}^n \phi_j \right\} \cL_k \left[ \exp \left\{i\sum_{\ell=1}^{k-1} \phi_{\ell} \right\}\tilde\beta \right] + \sum_{j,k}\Order \{ \eps_j\eps_k \} ,
\end{equation}
which follows directly by induction, with the convention $\sum_{n+1}^n(\cdots)\equiv 0$.

\subsection{Stroboscopic regime beyond weak signals}
\label{app:strobobeyondweak}

The stroboscopic regime also applies when the optomechanical coupling or mechanical signal is not weak. In this case, let us directly approximate $G_n(t) \approx G_n(0)$ in the time-domain recursion formula \eqref{eq:betan_recursion},
\begin{equation}
    \beta_n (t) \approx \beta_{n-1} (t) - \kappa_n  \int_{-\infty}^{\infty} \diff t' \, \Theta(t') e^{-G_n(0)t'} \beta_{n-1} (t-t'),
\end{equation}
with $\Theta$ is the Heaviside function.
Now we have a convolution that is conveniently solved in Fourier space. To this end, we introduce the Fourier transform of the convolution kernel and an associated signal phase,
\begin{align}
    S_n (\omega) &= \int\diff t \, \Theta(t) e^{-G_n(0)t+i\omega t} = \frac{1}{G_n (0) - i\omega} = \frac{1}{\Gamma_n - i[g_n Q_n(0)+\omega]}, \nonumber \\
    1-\kappa_n S_n (\omega) &=-\frac{\Gamma_n^* + i[g_n Q_n(0)+\omega] }{\Gamma_n - i[g_n Q_n(0) + \omega]} = -\frac{S_n (\omega)}{S_n^* (\omega)} =: e^{i\chi_n(\omega)},
    \label{eq:Sn}
\end{align}
\begin{equation}\label{eq:solution_betan_strobNotWeak}
    \tilde\beta_n(\omega) = \tilde\beta_{n-1}(\omega) - \kappa_n S_n (\omega) \tilde\beta_{n-1}(\omega) = e^{i\chi_n (\omega)} \tilde\beta_{n-1}(\omega)= \exp \left\{ i\sum_{k=1}^n \chi_k (\omega)\right\} \tilde\beta(\omega). 
\end{equation}
We also see an accumulating phase in the output spectral amplitude as the signature of coherent averaging.

\section{Thermal noise }\label{app:corrcovmatrix}
Inclusion of thermal and quantum back action noise requires a full quantum treatment, since, these two noise contributions arise when also the mechanical degree of freedom is treated as a quantum operator.  

The aim of this appendix is to show that both effects can be neglected under reasonable conditions. We restrict our considerations to the stroboscopic regime where the light pulse is much shorter than the mechanical period ($\tau^{-1}\gg \Omega$) while we still assume that it is much longer than the cavity decay time ($\kappa\gg \tau^{-1}$). Furthermore, we only consider the interaction of the light pulse with a single cavity.
This is reasonable since if we find that the contribution is small, then it can be added up linearly for many cavities.
Starting with the Hamiltonian in Eq.~\eqref{eq:hamiltonian}, we can drop the index $n$, and we need to add the free Hamiltonian for the mechanical oscillator $\hat H_M=\hbar \Omega \hat c^\dagger\hat c$, where $\oc^\dagger$ and $\oc$ are the creation and annihilation operator for the mechanical mode. 
The equations of motion are
\begin{equation}
\begin{cases}
&\partial_t \oc(t) = -i \Omega\oc(t)+i\dfrac{g}{\sqrt{2}}\oa^\dagger(t) \oa(t),\\
&\partial_t \oa(t)=-i\Delta \oa(t)+i \dfrac{g}{\sqrt{2}}\oa(t) (\oc^\dagger(t)+\oc(t))-\sqrt{\dfrac{\kappa}{2\pi}}\int \text{d} \omega\; \ob(\omega,t),\\
&\partial_t \ob(\omega,t)=-i(\omega-\omega_L) \ob(\omega,t)+\sqrt{\dfrac{\kappa}{2\pi}}\oa(t) .
\end{cases}
\end{equation} 
We assume that the resonator is initially in a thermal state at temperature $T$, which is then coherently displaced by the signal we aim to detect. For simplicity, we also assume that the thermalisation time is long compared to the probing time and the resonator state is thus given by a displaced thermal state oscillating at the mechanical frequency $\Omega $. What we measure is the modulation of the coherent output light amplitude due to this oscillation for a probe pulse of classical magnitude. Hence we can safely linearise the system of equations. Splitting each operator $\hat O$ into an expectation value $\bar O$ (here, $\alpha = \bar a$ and $\beta = \bar b$) and a zero-mean quantum fluctuation $\delta \hat O$ ($\langle \delta \hat O\rangle=0$) and omitting quadratic terms in the fluctuations, we obtain two sets of differential equations: one set for the quantum fluctuations,
\begin{equation}
\begin{cases}\label{eq:linearizedEOM}
&\partial_t \delta\oc(t) = -i \Omega\delta\oc(t)+i\dfrac{g}{\sqrt{2}}(\alpha^*(t)\delta\oa (t)+ \alpha(t)\delta\oa^\dagger(t)),\\
&\partial_t \delta\oa(t)=-i\Delta \delta\oa(t)+i \dfrac{g}{\sqrt{2}}\alpha(t) (\delta\oc^\dagger(t)+\delta\oc(t))+i\dfrac{g}{\sqrt{2}}\delta\oa(t)\left(\bar c^*(t)+\bar c(t)\right)-\sqrt{\dfrac{\kappa}{2\pi}}\int \text{d} \omega \delta\ob(\omega,t),\\
&\partial_t \delta \ob(\omega,t)=-i(\omega -\omega_L)\delta\ob(\omega,t)+\sqrt{\dfrac{\kappa}{2\pi}}\delta\oa(t) ;
\end{cases}
\end{equation}
and one for the classical averages,
\begin{equation}
\begin{cases}
&\partial_t \bar c(t) = -i \Omega\bar c(t)+i\dfrac{g}{\sqrt{2}}|\alpha(t)|^2,\\
&\partial_t \alpha(t)=-i\Delta \alpha(t)+i \dfrac{g}{\sqrt{2}}\alpha(t) (\bar c^*(t)+\bar c(t))-\sqrt{\dfrac{\kappa}{2\pi}}\int \text{d} \omega \beta(\omega,t),\\
&\partial_t \beta (\omega,t)=-i(\omega-\omega_L) \beta(\omega,t)+\sqrt{\dfrac{\kappa}{2\pi}}\alpha(t).
\end{cases}
\end{equation} 
Both systems of equations of motion can be solved.

\subsection{Solution for classical averages}

For the classical average, we take the formal solution of the input light field 
$\beta(\omega, t)=e^{-i(\omega-\omega_L)(t-t_0)}\beta(\omega,t_0)+\sqrt{\kappa/(2\pi)}\int_{t_0}^tdt' e^{-i(\omega-\omega_L)(t-t')}\alpha(t')$ and we substitute it in the differential equation for the cavity field $\alpha$ to get
\begin{equation}
    \partial_t\alpha(t)=\left[ -i\Delta+ig \bar{q}(t)-\dfrac{\kappa}{2}\right]\alpha(t)-\sqrt{\kappa}\beta(t)
\end{equation}
where $\bar{q}(t) =(\bar c^*(t)+\bar c(t))/\sqrt{2}$ is the mechanical position displacement and $\beta(t) = \sqrt{1/(2\pi)} \int d\omega\, e^{-i(\omega-\omega_L)(t-t_0)}\beta(\omega,t_0)$ corresponds to the input laser pulse centered around $t=0$, as defined in the main text.
The mechanical displacement is obtained by solving the equation for $\bar c(t)$, which splits into a term of zeroth order and a term of first order in the optomechanical strength $g$, 
\begin{equation}
    \bar c(t)=\bar c(t_0) e^{-i\Omega  (t-t_0)}+i\frac{g}{\sqrt{2}}\int_{t_0}^tdt' e^{-i\Omega (t-t')}|\alpha(t')|^2.
\end{equation}
We split the position displacement into the respective contributions, $\bar q(t)=\bar q^{(0)}(t)+\bar q^{(1)}(t)$, with
\begin{equation}
\begin{split}
    &\bar q^{(0)}(t)=\dfrac{\bar c(t_0)e^{-i\Omega (t-t_0)}+\bar c^*(t_0)e^{i\Omega (t-t_0)}}{\sqrt{2}},\\
    &\bar q^{(1)}(t)= g  \int_{t_0}^tdt' \sin(\Omega (t-t')) |\alpha(t')|^2 \approx g  \sin(\Omega t)\int_{t_0}^tdt'|\alpha(t')|^2. 
\end{split}\label{eq:mechCsol}
\end{equation}
The approximation holds in the stroboscopic regime in which the mechanical dynamics $\Omega$ is slow compared to the duration of the cavity response to the input laser pulse which leads to $|\alpha(t')|$ restricting significant contributions of the integrand to the regime $\Omega t'\ll 1$ around $t'=0$. Explicitly, the formal solution of the cavity amplitude is,
\begin{equation}
    \alpha(t)=\alpha(t_0)e^{-\Gamma(t-t_0)}e^{ig\int_{t_0}^tdt' \bar q(t')}-\sqrt{\kappa}\int_{t_0}^tdt'e^{-\Gamma(t-t')}e^{ig\int_{t'}^tdt'' \bar q(t'')}\beta(t'),
\end{equation}
abbreviating $\Gamma=\kappa/2+i\Delta$ as before. 
Since the first term vanishes for $t_0\rightarrow-\infty$, the amplitude is indeed non-vanishing only for the duration of the laser pulse. Moreover, since the pulse is slow in comparison to the cavity decay time, we can approximate $\beta(t') \approx \beta(t)$ in the second term. This also allows us to approximate $\int_{t'}^t dt'' \bar q(t'')\approx\bar q(0)(t-t')$ since the mechanical motion is slow compared to both the cavity decay and the pulse and the latter is centered around $t=0$. Noticing that $\bar{q}^{(1)}(0) \approx 0$, we arrive at the final expression for the cavity amplitude, 
\begin{equation}
    \alpha(t)\approx-\dfrac{\sqrt{\kappa}\beta(t)}{\Gamma-ig\bar q^{(0)}(0)}.
    \label{eq:classsolalpha}
\end{equation}

\subsection{Solution for the quantum fluctuations}
The linearised equations of motion \eqref{eq:linearizedEOM} for the quantum fluctuations can be solved analogously. We start with the solution for the light field,
\begin{equation}
    \delta \ob(\omega,t)=\delta \ob(\omega,t_0)e^{-i(\omega-\omega_L)(t-t_0)}+\sqrt{\dfrac{\kappa}{2\pi}}\int_{t_0}^tdt' e^{-i(\omega-\omega_L)(t-t')}\delta\oa(t),
\end{equation}
which we substitute into the equation for the cavity field fluctuations, 
\begin{equation}
    \partial_t \delta\oa(t)=-\left[\Gamma-ig\bar q(t)\right]\delta\oa(t)+ig\alpha(t)\delta\hat q(t)-\sqrt{\kappa}\delta\ob^{\rm (in)}(t),
\end{equation}
with $\delta\hat{q} (t) = [\delta\oc(t) + \delta \oc^\da (t)]/\sqrt{2}$.
Here, the input field operator describes vacuum noise on top of the coherent laser pulse, $\delta\ob^{\rm (in)}(t)=\sqrt{1/2\pi} \int d\omega e^{-i(\omega-\omega_L)(t-t_0)}\delta \ob(\omega,t_0)$.
The formal solution is
\begin{align}
    \delta\oa(t)&=\delta \oa(t_0) e^{-\int_{t_0}^tdt'(\Gamma-ig\bar q(t'))}+\int_{t_0}^tdt' e^{-\Gamma(t-t')+ig\int_{t'}^tdt'' \bar q(t'')}\left[ig\alpha(t')\delta\hat q(t')-\sqrt{\kappa} \delta\ob^{\rm (in)}(t')\right] \nonumber \\
    &\approx -\sqrt{\kappa} \int_{t_0}^tdt' e^{-[\Gamma-ig\bar{q}(t)](t-t')}\delta\ob^{\rm (in)}(t') + ig\alpha(t) \int_{t_0}^tdt' e^{-[\Gamma-ig\bar{q}^{(0)}(0)](t-t')}\delta\hat{q}(t') ,
    \label{eq:deltaasol}
\end{align}
which consists of a zeroth order and a first order contribution with respect to $g$, denoted by $\delta\oa^{(0)}(t)$ and $\delta\oa^{(1)}(t)$, respectively.
Here we exploited the fact that the first term in the first line vanishes for $t_0 \to -\infty$ due to the exponential decay with rate $\kappa/2$, and we made use of the approximations in the stroboscopic regime as before. In particular, $\alpha(t)$ is a Gaussian pulse centered around $t=0$ that is short compared to the mechanical oscillation time, but long compared to the cavity decay time. To proceed, we need the solution for the mechanical fluctuations,
\begin{equation}
    \delta\oc(t)=\delta \oc(t_0)e^{-i\Omega (t-t_0)}+i\dfrac{g}{\sqrt{2}}\int_{t_0}^tdt'e^{-i\Omega (t-t')}\left[\alpha(t')\delta\oa^\dagger(t')+\alpha^*(t')\delta \oa(t')\right],
\end{equation}
which again splits into a zeroth order and a first order contribution. In terms of the position fluctuations, $\delta\hat{q} (t) = \delta\hat{q}^{(0)} (t) + \delta\hat{q}^{(1)} (t)$, these contributions are, 
\begin{equation}
    \begin{split}
        \delta\hat q^{(0)}(t)&=\dfrac{\delta\oc(t_0)e^{-i\Omega (t-t_0)}+\delta\oc^\dagger(t_0)e^{i\Omega (t-t_0)}}{\sqrt{2}}, \\
        \delta\hat q^{(1)}(t)&= g\int_{t_0}^tdt' \sin\left[\Omega (t-t')\right]\left[\alpha^*(t')\delta\oa(t')+\alpha(t')\delta\oa^\dagger(t')\right]\\
        &\approx g\sin(\Omega t)\int_{t_0}^tdt'\left[\alpha^*(t')\delta\oa^{(0)}(t')+\alpha(t')\delta\oa^{(0)\dagger}(t')\right].
        \label{eq:deltaq1}
    \end{split}
\end{equation}
In the last line we used the stroboscopic approximation again, and we neglected second-order terms in $g$ by replacing the cavity fluctuation with its zeroth order contribution $\delta \oa^{(0)}$. Upon insertion into Eq.~\eqref{eq:deltaasol}, this amounts to neglecting terms of third order in $g$. Moreover, we notice that the $\delta\hat{q}^{(1)}$-contribution to $\delta\oa(t)$ vanishes effectively in the stroboscopic regime, since we can approximate $\sin(\Omega t') \approx \sin(\Omega t)$ under the integral and then  $\alpha(t)\sin (\Omega t) \sim O(\Omega \tau)$ is set to zero.
This contribution, which represents backaction noise, is thus negligible within the linearised stroboscopic regime. Introducing the abbreviation $G(t)=\Gamma-ig\bar q^{(0)}(t)$, we are left with
\begin{align}
    \label{eq:deltaa0}
        \delta\oa^{(0)}(t)&=-\sqrt{\kappa}\int_{t_0}^tdt' e^{-G(t)(t-t')}\delta\ob^{\rm (in)}(t'), \qquad \delta\oa^{(1)}(t)=ig \alpha(t)\int_{t_0}^tdt'e^{-G(0)(t-t')}\delta\hat q^{(0)}(t').
\end{align}

\subsection{Displacement vector and covariance matrix}

Assuming Gaussian statistics, the output field is fully determined by spectrally or time-resolved average displacements and by the respective covariances between the quantum fluctuations around those displacements. Expressed in terms of the time-domain quadrature vector, $\hat{\vd}(t)=\left(\ob^{\rm (out)}(t),\ob^{\rm (out)\dagger}(t)\right)^\intercal$, the time-resolved average displacements are simply given by the output amplitudes $\beta^{\rm (out)}(t)$. By virtue of the input-output relation, the latter comprise the input laser pulse and the cavity response, which is approximately given in \eqref{eq:classsolalpha}, so that
\begin{equation}\label{eq:displacementVector}
       \vd(t)=\la\hat{\vd}(t)\ra = \left(\beta(t)+\sqrt{\kappa}\alpha(t), \beta^{*}(t)+\sqrt{\kappa}\alpha^*(t)\right)^\intercal \approx -\left( \frac{G^*(0)}{G(0)}\beta(t), \frac{G(0)}{G^*(0)}\beta^{*}(t)\right)^\intercal.
\end{equation}
The time-resolved covariances are given by the matrix elements, 
\begin{equation}
 \sigma_{ab}(t_i,t_j)=\langle\{\hat d_a(t_i),\hat d^\dagger_b(t_j)\}\rangle-2\langle\hat{d}_a(t_i)\rangle\langle\hat{d}^\dagger_b(t_j)\rangle = \left\langle \left\{ \hat d_a(t_i)- d_a(t_i), \hat{d}^\dagger_b(t_j) - d^*_b(t_j) \right\} \right\rangle .
\end{equation}
Hence, they depend only on the quantum fluctuations of the output field modes. Explicitly, 
\begin{equation}
\begin{split}
    \sigma_{11}(t_i,t_j)&= \sigma_{22}(t_j,t_i) = \langle\delta\ob^{\rm (out)}(t_i)\delta\ob^{\rm (out)\da}(t_j)\rangle+\langle\delta\ob^{\rm (out)\da}(t_j)\delta\ob^{\rm (out)}(t_i)\rangle ,\\
    \sigma_{12}(t_i,t_j)&=  \sigma_{12}(t_j,t_i) = \sigma_{21}^*(t_i,t_j) = \langle\{\delta\ob^{\rm (out)}(t_i),\delta\ob^{\rm (out)}(t_j)\}\rangle .
   \end{split} 
   \label{eq:covmatrixFLUCT}
\end{equation}
By virtue of the input-output relation, we have
\begin{equation}
\delta \ob^{\rm (out)}(t)=\delta \ob^{\rm (in)}(t)+\sqrt{\kappa}\delta\oa^{(0)}(t)+\sqrt{\kappa}\delta\oa^{(1)} (t).
\end{equation}
Inserting the cavity terms from Eq.~\eqref{eq:deltaa0} allows us to expand the covariances in terms of first and second moments of the input field fluctuations and the unperturbed mechanical fluctuations. Most of these terms vanish, since $\langle\delta\ob^{\rm (in)\dagger}(t_i)\delta\ob^{\rm (in)}(t_j)\rangle=\langle\delta\ob^{\rm (in)\dagger}(t_i)\delta\ob^{\rm (in)\dagger}(t_j)\rangle=\langle\delta\ob^{\rm (in)}(t_i)\delta\ob^{\rm (in)}(t_j)\rangle=0$ and $\langle\delta \hat{q}^{(0)}(t)\rangle=0$. The only non-vanishing moments are either of zeroth order or of second order in $g$.
The zeroth order terms are:
\begin{align}
    \label{eq:IID}
    \langle\delta\ob^{\rm (in)}(t_j)\delta\ob^{\rm (in)\dagger}(t_i)\rangle&=\delta(t_i-t_j),\\
    \label{eq:I0D}
    \sqrt{\kappa}\langle\delta\ob^{\rm (in)}(t_j)\delta\oa^{(0)\dagger}(t_i)\rangle &= \sqrt{\kappa}\langle\delta\oa^{(0)}(t_i)\delta\ob^{\rm (in)\dagger}(t_j)\rangle^* = -\kappa e^{-G^*(t_i)(t_i-t_j)}\Theta(t_i-t_j),\\
    \kappa\langle\delta\oa^{(0)}(t_j)\delta\oa^{(0)\dagger}(t_i)\rangle&=\kappa^2\Theta(t_j-t_i) \dfrac{e^{-G(t_j)(t_j-t_i)}}{G(t_j)+G^*(t_i)}+\kappa^2\Theta(t_i-t_j)\dfrac{e^{-G^*(t_i)(t_i-t_j)}}{G(t_j)+G^*(t_i)} \nonumber \\
     \label{eq:00D}
    & \approx \kappa e^{-\kappa |t_j-t_i|/2-i(\Delta-g\bar{q}^{(0)}(t_i))(t_j-t_i)}
\end{align}
In the last line, we have exploited the stroboscopic regime once more: the fast exponential decay restricts the difference $|t_i-t_j|$ to small values of order $1/\kappa$ so that we can approximate $G(t_i)\approx G(t_j)$ and thus $G(t_j)+G^*(t_i) \approx \kappa$. Hence, in this regime, we find the following identity that will be important later: 
\begin{equation} \label{eq:dbda_identity}
    \sqrt{\kappa}\langle\delta\ob^{\rm (in)}(t_j)\delta\oa^{(0)\dagger}(t_i)\rangle + 
    \sqrt{\kappa}\langle\delta\oa^{(0)}(t_j)\delta\ob^{\rm (in)\dagger}(t_i)\rangle
    \approx - \kappa \langle\delta\oa^{(0)}(t_j)\delta\oa^{(0)\dagger }(t_i)\rangle .
\end{equation}

The second order terms are mixed second moments of $\delta \oa^{(1)}$ and its adjoint operator at two different times,
\begin{align}
    \label{eq:TDT}
    \kappa\langle\delta\oa^{(1)\dagger} (t_i)\delta\oa^{(1)} (t_j)\rangle &= \kappa g^2 \alpha^*(t_i)\alpha(t_j)\int_{t_0}^{t_i}dt_1e^{-G(0)(t_i-t_1)}\int_{t_0}^{t_j}dt_2e^{-G^*(0)(t_j-t_2)} \langle  \delta \hat q^{(0)}(t_1) \delta \hat q^{(0)}(t_2)\rangle,
   \\
   \label{eq:TTD}
 \kappa\langle\delta \oa^{(1)} (t_j)\delta \oa^{(1)\dagger} (t_i)\rangle&= \kappa g^2 \alpha(t_j)\alpha^*(t_i)\int_{t_0}^{t_j}dt_1e^{-G(0)(
   t_j-t_1)}\int_{t_0}^{t_i}dt_2e^{-G^*(0)(t_i-t_2)} \langle\delta \hat  q^{(0)}(t_1) \delta \hat q^{(0)}(t_2) \rangle, \\ 
   \label{eq:TT}
   \kappa\langle \delta \oa^{(1)} (t_i)\delta \oa^{(1)} (t_j)\rangle&=-\kappa g^2\alpha(t_i)\alpha(t_j)\int_{t_0}^{t_i}dt_1e^{-G(0)(t_i-t_1)}\int_{t_0}^{t_j}dt_2e^{-G(0)(t_j-t_2)}\langle\delta \hat q^{(0)}(t_1)\delta\hat q^{(0)}(t_2)\rangle.
\end{align}
They all depend on two-time correlations of the mechanical position fluctuations, $\langle\delta \hat q^{(0)}(t_1)\delta \hat q^{(0)}(t_2)\rangle$, with respect to the displaced thermal state we assume for the mechanical mode. These are given by the respective correlations of the full position operators with respect to the undisplaced thermal (Gibbs) state, 
\begin{equation}\label{eq:thermal_fluc}
    \langle\delta \hat q^{(0)}(t_1)\delta \hat q^{(0)}(t_2)\rangle = \Tr[\rho_{\rm th}\hat q^{(0)}(t_1)\hat q^{(0)}(t_2)] = \bar{n}\cos\Omega (t_1-t_2)+\dfrac{1}{2}e^{-i\Omega (t_1-t_2)},
\end{equation}
with $\bar{n} = [\exp(\hbar\Omega /k_{\rm B} T)-1]^{-1}$ the thermal occupation number. As the correlations oscillate with the slow mechanical frequency, we can invoke the stroboscopic regime again and approximate their time arguments by $(t_1,t_2) \approx (0,0)$ due to the fast cavity decay and the short pulse duration in Eqs.~\eqref{eq:TDT}-\eqref{eq:TT}. Hence,
\begin{align}
    \kappa\langle\delta\oa^{(1)\dagger} (t_i)\delta\oa^{(1)} (t_j)\rangle &\approx \frac{\kappa g^2}{|G(0)|^2} \alpha^*(t_i)\alpha(t_j) \frac{2\bar{n}+1}{2} \approx \kappa\langle\delta \oa^{(1)} (t_j)\delta \oa^{(1)\dagger} (t_i)\rangle, \\
    \kappa\langle \delta \oa^{(1)} (t_i)\delta \oa^{(1)} (t_j)\rangle &\approx -\frac{\kappa g^2}{G(0)^2} \alpha(t_i)\alpha(t_j) \frac{2\bar{n}+1}{2}.
\end{align}
We can now gather all the non-vanishing moments to obtain expressions for the covariance matrix elements in \eqref{eq:covmatrixFLUCT}. For $\sigma_{11}$, we get
\begin{align}\label{eq:sigma11}
    \sigma_{11} (t_i,t_j) &=  
    \langle\delta\ob^{\rm (in)}(t_i)\delta\ob^{\rm (in)\dagger}(t_j)\rangle + \kappa \left[ \langle\delta\oa^{(1)\dagger}(t_j)\delta\oa^{(1)}(t_i)\rangle+\langle\delta\oa^{(1)}(t_i)\delta\oa^{(1)\dagger}(t_j)\rangle \right] \nonumber \\
    &\quad +\sqrt{\kappa} \langle\delta\oa^{(0)}(t_i)\delta\ob^{\rm (in)\dagger}(t_j)\rangle +  \sqrt{\kappa} \langle\delta\ob^{\rm (in)}(t_i)\delta\oa^{(0)\dagger}(t_j)\rangle + \kappa \langle\delta\oa^{(0)}(t_i)\delta\oa^{(0)\dagger}(t_j)\rangle \nonumber \\
    &\approx \delta (t_i - t_j) + (2\bar{n}+1)\frac{\kappa g^2 }{|G(0)|^2} \alpha^*(t_j)\alpha(t_i) \approx \delta (t_i - t_j) + (2\bar{n}+1)\frac{\kappa^2 g^2 }{|\Gamma|^4} \beta^{*}(t_j)\beta(t_i).
\end{align}
Here, we made use of Eq.~\eqref{eq:dbda_identity} to cancel the second line, we inserted the approximate expression \eqref{eq:classsolalpha} for the cavity amplitude, and we consistently omitted terms higher than second order in $g$ by replacing $G(0)$ with $\Gamma$. 
Similarly, for $\sigma_{12}$ we obtain 
\begin{align}\label{eq:sigma12}
    \sigma_{12} (t_i,t_j) &= 
    \kappa \langle\{\delta\oa^{(1)}(t_i),\delta\oa^{(1)}(t_j)\}\rangle \approx -(2\bar{n}+1)\frac{\kappa g^2 }{G(0)^2} \alpha(t_j)\alpha(t_i) \approx -(2\bar{n}+1)\frac{\kappa^2 g^2 }{\Gamma^4} \beta (t_j)\beta(t_i) .
\end{align}
The remaining two matrix elements $\sigma_{22}$ and $\sigma_{21}$ are related to the given ones through Eq.~\eqref{eq:covmatrixFLUCT}. 
Overall, the thermal noise contributes a rank-1 update of the covariance matrix, $\vsigma = \vsigma_0+\vw \vw^\da$, as given by the quadrature vector $\vw = \kappa g \sqrt{2\bar n+1} (i\vbeta/\Gamma^2,-i\vbeta^*/\Gamma^{*2})^\intercal$. Here, $\vbeta$ is the vector with $i$-th component given by $\beta(t_i)$ and $\vw\vw^\da$ denotes the complex outer product. Explicitly in the time domain, the vector elements are $w_1(t_i) = i\kappa g \sqrt{2\bar n+1} \beta(t_i)/\Gamma^2$ and $w_2(t_i) = w_1^*(t_i)$, so that one arrives at the matrix elements $\sigma_{ab} (t_i,t_j)$ in \eqref{eq:sigma11} and \eqref{eq:sigma12}.
In the limit $g\to 0$, the covariance matrix reduces to the trivial one for the free mode continuum, $\vsigma \to \vsigma_0 = \id$, as expected.

\subsection{Quantum Fisher Information}
To evaluate the QFI with respect to the signal strength parameter $\vartheta$, as defined in Eq.~\eqref{eq:QFIdef}, the inverse of the covariance matrix is required. 
Since the thermal noise correction is a rank-1 update of the free covariance matrix in terms of the above defined vector $\vw$, we can apply the complex-valued version of the Sherman-Morrison formula for the matrix inverse \cite{bartlett1951}, 
\begin{equation}
    \vsigma^{-1} = \id - \frac{\vw\vw^\da}{1+\vw^\da \vw} = \id - \frac{\vw\vw^\da}{1+2\frac{\kappa^2g^2}{|\Gamma|^4}(2\bar n + 1) N_{\rm in}},
\end{equation}
with the number of input photons $N_{\rm in} = \int \diff t |\beta(t)|^2 = \sqrt{\pi}\tau \bar\beta^2$. The term in the denominator displays the strength of the thermal noise correction; it is negligible for arbitrary detuning $\Delta$ as long as $32(g/\kappa)^2 (2\bar n+1)N_{\rm in} \ll 1$, in which case its effect for $N$ oscillators can be treated as additive and thus the overall enhancement is only marginally affected. The QFI \eqref{eq:QFIdef} becomes,
\begin{equation}
   \mathcal{H}= 2\int dt\, |\partial_\vartheta\vd(t)|^2-2 \frac{\left|\int \diff t\,  \vw^\da(t)\partial_\vartheta\vd (t)\right|^2}{1+2\frac{\kappa^2g^2}{|\Gamma|^4}(2\bar n + 1) N_{\rm in}} =: \mathcal{H}_0+ \delta\mathcal{H}, 
\end{equation}
where the first term $\mathcal{H}_0$ corresponds to the QFI in the absence of thermal and back-action noise and the second term $\delta \mathcal{H}$ represents the correction due to thermal noise, as detailed in the previous subsection (showing also that back action noise is negligible). Notice that there would be an additional contribution to the QFI coming from the $\vartheta$-dependence of the covariance matrix. However, we showed that this dependence is of higher than second order in the optomechanical coupling $g$ and can therefore omit it.

The signal strength parameter $\vartheta$ thus enters the QFI only through the derivative of the displacement vector \eqref{eq:displacementVector}. Assuming, as in the main text, that $\vartheta$ describes the signal amplitude up to a proportionality constant, we set the mechanical position displacement at $t=0$ to $\bar{q}(0) = Q\vartheta$ and arrive at
\begin{equation}
   \partial_\vartheta \vd(t) \approx \left(-\frac{ig\kappa Q}{\Gamma^2}\beta(t), 
   \frac{ig\kappa Q}{\Gamma^{*2}}\beta^{*}(t)\right)^\intercal, 
\end{equation}
dropping higher order contributions of $g$ again, consistently. 
The free-field QFI without thermal noise simplifies to
\begin{equation}
   \mathcal{H}_0=\dfrac{4 g^2\kappa^2 Q^2 }{|\Gamma|^4} N_{\rm in} = \frac{ 64 g^2\kappa^2 Q^2}{(\kappa^2 + 4\Delta^2)^2} N_{\rm in},
   \label{eq:Hfree}
\end{equation}
For the noise correction of the QFI, we are left with 
\begin{equation}
    \delta \mathcal{H} = -  \frac{8g^4\kappa^4 Q^2(2\bar n + 1)N_{\rm in}^2}{|\Gamma|^8+2\kappa^2g^2|\Gamma|^4(2\bar n + 1) N_{\rm in}} = -\frac{(2\bar{n}+1)\mathcal{H}_0}{2Q^2+(2\bar n + 1) \mathcal{H}_0} \mathcal{H}_0 
\end{equation}
The correction is always negative, as the thermal noise deteriorates coherence and thus reduces phase sensitivity. In the limit of high noise-free sensitivity, $(2\bar n + 1)\mathcal{H}_0 \gg 2Q^2$, the thermal noise would diminish the QFI almost completely as $\delta \mathcal{H} \to -\mathcal{H}_0$. On the other hand, in the here relevant weak-signal regime, $(2\bar n + 1)\mathcal{H}_0 \ll 2Q^2$, 
we can distinguish two limiting cases. At low temperatures, $k_{\rm B} T \ll \hbar\Omega$, we have $\bar n \approx 0$, and the mechanical zero-point fluctuations diminish the QFI by the small constant fraction $|\delta \mathcal{H}/\mathcal{H}_0| \approx \mathcal{H}_0/2 Q^2{\ll 1}$. More realistically, for low oscillator frequencies and comparably high temperatures, $k_{\rm B} T \gg \hbar\Omega$, we have $2\bar n + 1 \approx 2k_{\rm B} T / \hbar\Omega$, and the relative correction is $|\delta \mathcal{H}/\mathcal{H}_0| \approx \mathcal{H}_0 k_{\rm B} T/ \hbar\Omega Q^2$.

This gives a condition under which we can safely neglect the contributions of thermal fluctuations of the mechanical element to the covariance matrix of the output light field. It is 
\begin{equation}\label{eq:tempcond}
    T\ll T_{\rm max}=\dfrac{Q^2\hbar\Omega }{k_{\rm B}\mathcal H_0}.
\end{equation}
We consider typical optomechanical parameters, namely  $\kappa/2\pi\approx  10^6~\mathrm{Hz}$, $\Omega/2\pi\approx  10^3~\mathrm{Hz}$, $\Delta\approx\Omega$. Working in the weak-coupling regime with $g/2\pi\approx 1~\mathrm{Hz}$ and taking $N_{\rm in}\approx100$, we further set the mechanical signal parameters to $Q=1$ and $\varphi=0$ without loss of generality. For this choice of parameters, we obtain  $T_{\rm max}\approx 7.5~\mathrm{K}$. In this case, $T\ll T_{\rm max}$ can be achieved with simple cooling techniques.

\section{Checks of pulse length approximations}
\label{app:checksofapprox}

We aim to check the validity of the stroboscopic and continuous-wave approximations made in the main text, which are standard approximations in the literature. To this end, we focus on the case of a single optomechanical cavity illuminated by a Gaussian light pulse, $\beta_0(t) = \bar\beta e^{-t^2/2\tau^2}$. Assuming that we aim to estimate the amplitude of the mechanical signal, $Q(t) \propto \vartheta$, the relevant quantity (excluding corrections to the covariance matrix) that determines the QFI is
\begin{equation}\label{eq:L_L2norm}
    \mathcal{I}:=\int\diff \omega \left|\partial_{\vartheta} \cL[\tilde\beta_0] (\omega) \right|^2 = \frac{1}{\vartheta^2} \| \cL[\tilde\beta_0] \|_2^2 = \frac{1}{\vartheta^2} \int\diff t \, Q^2(t) |R(t)|^2 , \qquad \text{with}\quad R(t)=\int \frac{\diff \omega}{\sqrt{2\pi}} \left[1-e^{i\phi(\omega)} \right]^2\tilde\beta_0(\omega) e^{-i \omega t} .
\end{equation}
Here, we have made use of Parseval's theorem and the explicit form of the integral kernel \eqref{eq:Ln_main}. Given the Fourier transform of the cavity response term,
\begin{equation}
  r(t) = \int \frac{\diff \omega}{\sqrt{2\pi}}  \left[1-e^{i\phi(\omega)} \right]^2 e^{-i\omega t} = \int \frac{\diff \omega}{\sqrt{2\pi}} \frac{\kappa^2 e^{-i\omega t}}{(\Gamma-i\omega)^2}=\sqrt{2\pi}\kappa^2 t \Theta(t) e^{-\Gamma t}, 
\end{equation}
we can simplify
\begin{equation}
    R(t) = \int \frac{\diff t'}{\sqrt{2\pi}} r(t') \beta_0 (t-t') = \kappa^2 \int_0^\infty \diff t' \, t' e^{-\Gamma t'} \beta_0 (t-t') = \kappa^2 \tau^2 \beta_0 (t) + \sqrt{2\pi}\kappa^2 \tau \bar\beta  \frac{t-\Gamma \tau^2}{2}\left[1+ \mathrm{erf} \left( \frac{t-\Gamma\tau^2}{\sqrt{2}\tau} \right) \right] e^{\Gamma^2\tau^2/2-\Gamma t} .
\end{equation}
This admits a convenient numerical evaluation of \eqref{eq:L_L2norm} and comparison to approximate expressions. 

\subsection{Stroboscopic regime}

\begin{figure}
    \centering
    \includegraphics[width=0.5\linewidth]{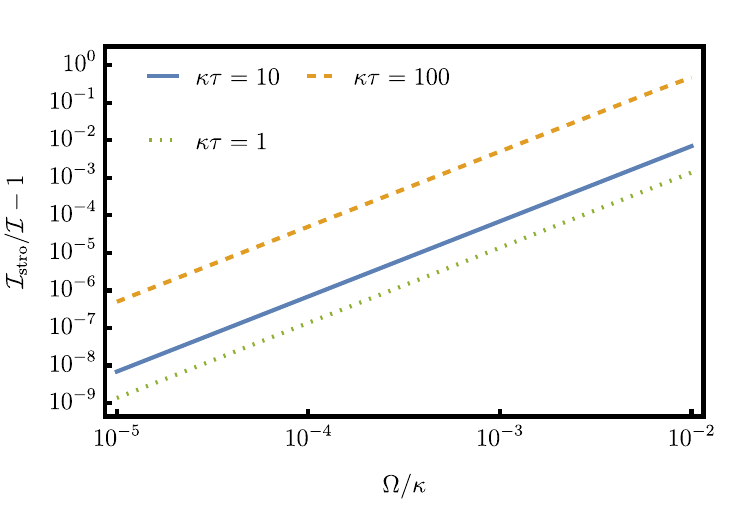}
    \caption{Relative deviation of the stroboscopic approximation for $\mathcal{I}$ from the full numerical result plotted over the mechanical frequency in units of $\kappa$ for different values of $\tau\kappa$.}
    \label{fig:numerical-test_strobo}
\end{figure}

For the case of the stroboscopic regime, we consider a continuous mechanical signal with $\Omega\ll 1/\tau$, write $Q(t)=\vartheta\cos(\Omega t)$, and compare the full QFI integral $\mathcal{I}$ with the stroboscopic approximation
\begin{equation}
    \mathcal{I}_{\rm strob}:= \frac{1}{\vartheta^2} Q^2(0) \int\diff t \,  |R(t)|^2 . 
\end{equation}
The numerical results for the case $\Delta =0$ (which should be optimal for the stroboscopic regime) are presented in Fig.~\ref{fig:numerical-test_strobo}. The error is negligible in the full plotted regime. Expanding $Q(t)^2\approx \vartheta^2 (1 - \Omega^2t^2)$ to second order in $\Omega t$ around the center of the pulse at $t=0$, we find that the error scales with $(\Omega \tau)^2$ for long pulses as 
\begin{equation}
    \mathcal{I} \approx \mathcal{I}_{\rm strob} - \Omega^2\int\diff t \, t^2  |R(t)|^2 
\end{equation}
in this regime.

\subsection{Continuous-wave regime}

\begin{figure}
    \centering
    \includegraphics[width=0.5\linewidth]{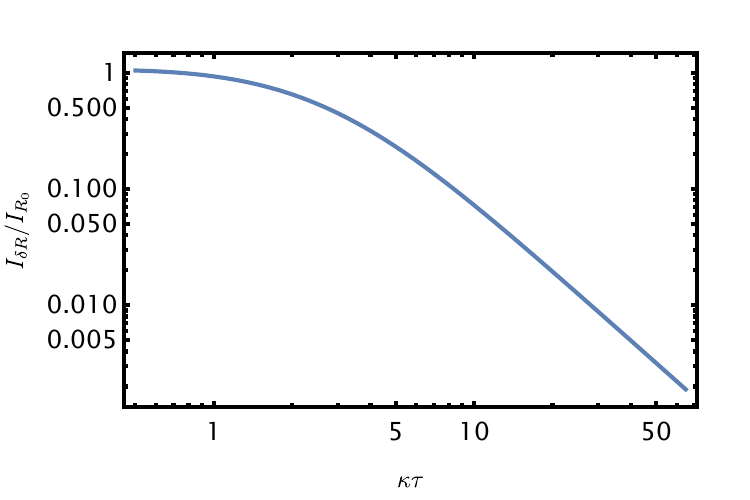}
    \caption{Ratio of the square integrated error in the response function for the continuous wave approximation divided by the square integrated approximate expression of the response function which is valid for $\kappa\tau\gg 1$.}
    \label{fig:numerical-test_continous_1}
\end{figure}
In the continuous wave regime, $\Omega\tau\gg 1$, we use a finite mechanical signal pulse with a pulse width in time $\sigma=1/\Delta\Omega \ll \tau$. In the following, we compare $R(t)$ with the corresponding result $R_0(t)$ that we obtain by assuming $\kappa \tau\gg 1$ which corresponds to the pulse being spectrally much narrower than the cavity response, i.e.,
\begin{equation}
    R_0(t)= \int \frac{\diff \omega}{\sqrt{2\pi}} \frac{\kappa^2 e^{-i\omega t}}{(\Gamma-i\omega)^2}\tilde\beta_0(\omega) \approx  \frac{\kappa^2}{\Gamma^2}\beta_0(t)
\end{equation}
In the resonant case, $\Delta=0$, the approximated cavity response function for $\kappa\tau\gg 1$ simplifies to $R_0(t)=4\beta_0(t)$. We check the size of the difference $\delta R=R-R_0$ square-integrated over $t$ 
\begin{equation}
    I_{dR}=\int dt |\delta R(t)|^2
\end{equation}
relative to $R_0$ square integrated over $t$, i.e.,
\begin{equation}
    I_{R_0} = \int dt |R_0|^2 = 16\sqrt{\pi} \tau\,.
\end{equation}
The result in shown in Fig.~\ref{fig:numerical-test_continous_1} and shows that the relative error is of the order of $10^{-3}$ or lower for $\kappa\tau\gtrsim 50$. 

\begin{figure}
    \centering
    \includegraphics[width=0.5\linewidth]{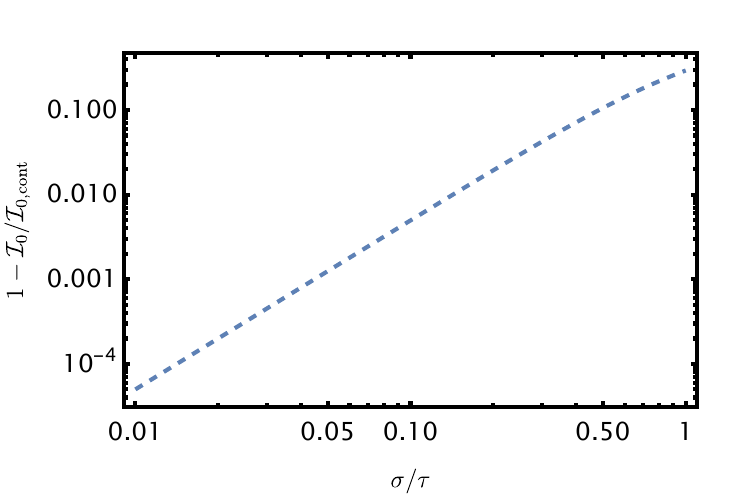}
    \caption{Ratio of the relative error of the CW approximation of the main text as a function of $\sigma/\tau$.}
    \label{fig:numerical-test_continous_2}
\end{figure}
Now we use $R_0$ to calculate $\mathcal{I}$ and a Gaussian mechanical signal
\begin{equation}
    Q(t)=\vartheta \cos(\Omega t) e^{-\frac{t^2}{2\sigma^2}}\,.
\end{equation}
We stick to the case $\Delta=0$. The integral has an analytical solution 
\begin{equation}
    \mathcal{I}_0 := \frac{1}{\vartheta^2} \int\diff t \,  |Q(t)R_0(t)|^2 = \frac{8 \sqrt{\pi } \left(e^{-\frac{\sigma ^2 \tau ^2 \Omega ^2}{\sigma ^2+\tau ^2}}+1\right)}{\sqrt{\frac{1}{\sigma ^2}+\frac{1}{\tau ^2}}}
\end{equation}
which we compare to the CW approximation we made in the main text, i.e.,
\begin{equation}
    \mathcal{I}_{0,{\rm cont}} := \frac{1}{\vartheta^2} |R_0(0)|^2 \int\diff t \,  |Q^2(t)|^2 = 8 \sqrt{\pi } \sigma  \left(e^{-\sigma ^2 \Omega ^2}+1\right).
\end{equation}
Expanding the difference of the two functions to third order in $\sigma/\tau$ leads to
\begin{equation}
    1-\mathcal{I}_0/\mathcal{I}_{0,{\rm cont}} \approx \left(\frac{\sigma}{\tau}\right)^2 \left(e^{-\sigma ^2 \Omega ^2}+1\right)^{-1}
\end{equation}
which decays quadratically with decreasing $\sigma/\tau$. A plot of the relative error is shown in Fig.~\ref{fig:numerical-test_continous_2}.

\subsection{Continuous-wave regime for continuous signals}

In this regime, we assume that $\Omega\tau\gg 1$, $\kappa\tau\gg 1$, and in addition, $\Delta\Omega \tau \ll 1$. We consider the monochromatic signal $Q(t)=\vartheta \cos(\Omega t)$ and the response function $R_0(t)$ which we have justified above for $\kappa\tau\gg 1$. Then, we find for the QFI integral taking into account the explicit form of $\beta_0(t)$
\begin{equation}\label{eq:Icontsig}
\mathcal{I}_{\mathrm{cont.sig.}}
:=
\frac{1}{\vartheta^2}
\int dt\, |Q(t)R_0(t)|^2
= \frac{\kappa^4}{|\Gamma|^4} \bar{\beta}^{\,2} \frac{\sqrt{\pi}\tau}{2} \left(1+e^{-\Omega^2\tau^2}\right).
\end{equation}
Alternatively, we can start directly from $\partial_{\vartheta} \cL[\tilde\beta_0] (\omega)$. In the described regime, we find
\begin{equation}
    \tilde Q(\omega)=\vartheta \frac{\sqrt{2\pi}}{2} \left(\delta(\omega-\Omega) + \delta(\omega+\Omega)\right).
\end{equation}
Therefore,
\begin{equation}
    \cL[\tilde\beta_0] (\omega) = \frac{\vartheta}{2}\left( \left[1-e^{i\phi(\omega-\Omega)} \right]^2\tilde\beta_0(\omega-\Omega)  + \left[1-e^{i\phi(\omega+\Omega)} \right]^2\tilde\beta_0(\omega+\Omega)  \right).
\end{equation}
Using the assumptions $\Omega\tau\gg 1$ and $\kappa\tau\gg 1$, the pulse spectrum $\tilde\beta_0(\omega)$ is sharply localized around $\omega=0$ in comparison to the cavity response and we can evaluate $\phi(\omega\pm\Omega)$ at $\omega\pm\Omega=0$ in the expression above such that
\begin{equation}
    \cL[\tilde\beta_0] (\omega) \approx \frac{\vartheta\kappa^2}{2\Gamma^2}\left(\tilde\beta_0(\omega-\Omega)  + \tilde\beta_0(\omega+\Omega)  \right).
\end{equation}
which corresponds to the side-band structure of the signal. Eventually, we recover the right hand side of Eq.~\eqref{eq:Icontsig}, where the term $e^{-\Omega^2\tau^2}$ can now be identified as resulting from the finite overlap of the two sidebands. Since we assume $\Omega\tau\gg1$, it is justified to neglect the overlap as stated in the main text. The relative error will decrease exponentially with $(\Omega\tau)^2$.

\twocolumngrid
\bibliography{references}

\end{document}